\documentclass[aps,showpacs,preprintnumbers,amsmath,amssymb,nofootinbib,eqsecnum, preprintnumbers]{revtex4}
\usepackage{graphicx}
\usepackage{epsf}
\usepackage{amsmath}
\usepackage{epstopdf}
\usepackage{bm}
\usepackage{color}
\usepackage{tabularx}
\usepackage{enumitem}
\usepackage{float}
\usepackage{array,booktabs}
\usepackage{footnote}
\usepackage{threeparttable}
\usepackage{graphicx}
\usepackage{hyperref}
\usepackage{amssymb,epsf}
\usepackage{latexsym}
\usepackage{epstopdf}
\usepackage{epsfig}
\usepackage{mathtools}

\begin{document}

\title{Holographic Heat Engines Coupled with Logarithmic $U(1)$ Gauge Theory}
\author{Soodeh Zarepour$^{1}$\footnote{email address: szarepour@phys.usb.ac.ir}} \affiliation{$^1$ Department of Physics, University of Sistan and Baluchestan, Zahedan, Iran}

\begin{abstract}
In this paper we study a new class of holographic heat engines via charged AdS black hole solutions of Einstein gravity coupled with logarithmic nonlinear $U(1)$ gauge theory.  So, Logarithmic $U(1)$ AdS black holes with a horizon of positive, zero and negative constant curvatures are considered as a working substance of a holographic heat engine and the corrections to the usual Maxwell field are controlled by non-linearity parameter $\beta$. The efficiency of an ideal cycle ($\eta$), consisting of a sequence of isobaric $\to$ isochoric $\to$ isobaric $\to$ isochoric processes, is computed using the exact efficiency formula. It is shown that $\eta/\eta_{C}$, with $\eta_{C}$ the Carnot efficiency (the maximum efficiency available between two fixed temperatures), decreases as we move from the strong coupling regime ($\beta \to 0$) to the weak coupling domain ($\beta \to \infty$). We also obtain analytic relations for the efficiency in the weak and strong coupling regimes in both low and high temperature limits. The efficiency for planar and hyperbolic logarithmic $U(1)$ AdS black holes is computed and it is observed that efficiency versus $\beta$ behaves in the same qualitative manner as the spherical black holes.

\end{abstract}

 \maketitle

\section{Introduction}
 The gauge/gravity duality (also known as AdS/CFT correspondence) relates the physics of a gravitational theory in an asymptotically anti de Sitter (AdS) background to the physics of a conformal field theory (CFT) at the AdS boundary \cite{Maldacena1998}. This elegant discovery is the primary motivation for studying black holes in AdS space \cite{Witten1998a,Witten1998b,PhysicsReports2000}. In the most recognized version of this duality, $\rm{AdS}_5/\rm{CFT}_4$ correspondence, the parameters of the supergravity theory and the CFT satisfy the following dictionary \cite{Maldacena1998,NatsuumeBookAdS/CFT}
  \begin{equation}
 \frac{{{L^3}}}{{{G_5}}} = \frac{{2N_c^2}}{\pi },\,\,\,\,\lambda  = {\left( {\frac{L}{{{\ell _s}}}} \right)^4},
  \end{equation}
 where $L$ is the AdS radius, $G_5$ is the Newton's constant, $\lambda$ is the $^{,}$t Hooft coupling, $\ell_s$ is the string length, and $N_c$ is the number of the colors. In the gravitational theory and naturally in gauge/gravity duality, the cosmological constant ($\Lambda$), which relates to the AdS radius by $\Lambda \propto 1/{L^2}$, is an external fixed parameter. However, the recent and interesting development in black hole physics suggests that the mass of AdS black holes should be interpreted as the enthalpy via extending the phase space, i.e., treating the cosmological constant as a thermodynamic variable (pressure) \cite{Kastor2009,Kastor2010,Dolan2011a,Dolan2011b,Kubiznak2012}. Taking this proposal into account, the variation of $\Lambda$ in the gravity side is equivalent to the variation of the number of colors ($N_c$) in the field theory side, provided that the string coupling ($g_s$) is held fixed \cite{Kastor2009}. This leads to a new holographic interpretation of AdS black holes since the dynamics of the CFT crucially depends on the number of colors, and especially, the holographic origin of the Smarr relation is explained \cite{Karch2015}. 
\par
Variation of physical constants leads to considering more fundamental theories in physics. Of interesting case is the use of the cosmological constant as a dynamical variable in black hole physics (see the earlier works in \cite{EarlierWorksA,EarlierWorksB,EarlierWorksC}). Along this line, extending the thermodynamic phase space of black holes by identifying the varying cosmological constant as a thermodynamic pressure, $P=-\Lambda/8\pi $, has opened up new theoretical avenues. In fact, the cosmological constant ($\Lambda$) has a privileged role to be identified as the thermodynamic pressure. There are several strong reasons for supporting this claim:

\begin{itemize}
  
\item Besides that $\Lambda$ generally appears as pressure in cosmology \cite{Weinberg2008Cosmology}, it has dimension of pressure in black hole thermodynamics as well (i.e., $[length]^{-2}$ in geometric units,  
${G_N} = \hbar  = c = {k_{\rm{B}}} = 1$) and also its conjugate variable has dimensions of volume \cite{Kastor2009,Kubiznak2012}.

\item Having this identification, one can naturally obtain the conjugate volume using the standard thermodynamic relation as $V=\partial M/\partial P$ \cite{Kastor2009,Kubiznak2012}. This definition avoids the difficulties of the other definitions of black hole volume \cite{Parikh2006,VectorVolume2013,Rovelli2015}, in which the volume integration has to be performed inside the horizon and this is problematic in some ways since the interior of black hole metric is not static.

\item Remarkably, this thermodynamic volume ($V=\partial M/\partial P$) is supported by the geometric definition of black hole volume by means of the Komar integral relation as (found by Kastor et al \cite{Kastor2009})
\begin{equation}
V =  - \left[ {\int_{\partial {\Sigma _\infty }} {d{S_{ab}}\left( {{\omega ^{ab}} - \omega _{{\rm{AdS}}}^{ab}} \right) - \int_{\partial {\Sigma _{\rm{h}}}} {d{S_{ab}}{\omega ^{ab}}} } } \right],
\end{equation}
where $d{S_{ab}}$ is the volume element normal to the co-dimension 2 surface $\partial \Sigma $ and the Killing potential ($\omega ^{ab}$) is precisely defined by use of Killing vector as ${\xi ^a} = {\nabla _b}{\omega ^{ab}}$. It is an interesting finding since this quantity gives a measure of the volume excluded from the spacetime by the black hole's horizon, as well understood in \cite{Kastor2009}, so it is interpreted as an effective volume inside the event horizon.
  
\item On the other hand, when a cosmological constant is present, the first law of black hole thermodynamics in the traditional treatments becomes inconsistent with the Smarr relation which means that the scaling argument (Euler’s theorem in thermodynamics) is no longer valid. This difficulty can be cured by identifying $\Lambda$ proportional to the thermodynamic pressure \cite{Kastor2009,Kubiznak2012,Review2017}.

\item There exist further pieces of evidence which show that black hole thermodynamics has a self-consistent framework within the extended phase space (for a nice review see \cite{Review2017}). For example, in the extended phase space, the Ehrenfest equations have well-defined definitions for a genuine second-order phase transitions \cite{Ehrenfest2013,Ehrenfest2014}. As another example, comparing the Van der Waals phase transition in the extended phase space of AdS black holes and also in the Van der Waals fluids shows a precise analogy between physical quantities of AdS black holes and Van der Waals fluids \cite{Kubiznak2012}. But, the similarity between the black hole thermodynamics and everyday thermodynamics found in the non-extended phase spaces (e.g., see Refs. \cite{Banerjee2010,Banerjee2011,Chamblin1999a,Chamblin1999b}) are some mathematical analogies rather than exact correspondences, as indicated in \cite{Kubiznak2012,Review2017}. However, in the extended phase space, the analogies are more direct \cite{Kubiznak2012,Ehrenfest2013,Ehrenfest2014,Review2017}.
\end{itemize}
\par
All these results seem to justify that the varying cosmological constant as thermodynamic pressure has a privileged role. In conclusion, this identification leads to a number of interesting results, most importantly, the first law of black hole thermodynamics is defined the same as everyday thermodynamics. Consequently, one can think about the other branches of thermodynamics and apply them in black holes as well. Pushing on this idea, interesting phase transitions have emerged like van der Waals phase structure in a number of AdS black hole spacetimes \cite{Kubiznak2012,Mann2012,Review2017,vdWNED2019}. Another novel direction is studying of black holes as heat engines \cite{Johnson2014} and exploring various thermodynamic cycles in them which has attracted a great deal of attention lately \cite{HeatEngines2015a,HeatEngines2015b,Johnson2016a,Johnson2016b,HeatEngines2017a,HeatEngines2017b,HeatEngines2017c,HeatEngines2017d,HeatEngines2017e,Johnson2018a,Johnson2018b,Johnson2018c,HeatEngines2018a,HeatEngines2018b,HeatEngines2018c,HeatEngines2018d,HeatEngines2018e,Johnson2019,HeatEngines2019a,HeatEngines2019b,HeatEngines2019c,HeatEngines2019d,HeatEngines2019e,HeatEngines2020}. In what follows, we shall restrict ourselves to this subject.
 \par
 Considering the extended phase space, it is natural to use black holes as heat engines \cite{Johnson2014} and explore how different closed loops in the $P-V$ plane can be realized as thermodynamic cycles. A number of interesting results have been found for various classes of AdS black holes. Born-Infeld AdS black holes have been considered as the working substance in Ref. \cite{Johnson2016a}, proving that the engine's efficiency is affected by the non-linearity parameter, $\beta$. A generalization of this study has been performed via Einstein-dilaton-Born-Infeld system in Ref. \cite{HeatEngines2017e}. Corrections of higher order curvatures to heat engines is also examined in Ref. \cite{Johnson2016b}, where, from the perspective of the dual holographic large $N_{c}$ field theory, it amounts to studying the effects of a class of $1/N_{c}$ corrections to the efficiency of the engine's cycle. Holographic heat engines have also been constructed in the context of massive gravity in Refs. \cite{HeatEngines2018a,HeatEngines2018b} and it is shown that the existence of graviton mass could improve the heat engine efficiency significantly. These types of holographic heat engines demonstrate that corrections arisen from gravity model or matter fields affect the engine's performance and efficiency. As an another example, properties of holographic heat engines with charged accelerating AdS black holes as the working substance is investigated in a benchmarking scheme \cite{HeatEngines2018c, HeatEngines2019a}. The heat engines' properties are considered for static AdS black holes in higher and lower dimensions \cite{HeatEngines2015b,HeatEngines2017c,HeatEngines2019d}, as well as in other black hole backgrounds such as Taub-Bolt solutions, rotating AdS solutions etc \cite{HeatEngines2015a,HeatEngines2017b,HeatEngines2017d,Johnson2018a,HeatEngines2018d,HeatEngines2018e,HeatEngines2019b,HeatEngines2020}. There have been several papers on AdS black holes as heat engines which can be viewed as a toy model for further investigation of holographic heat engines which opened new avenues \cite{HeatEngines2017a,Johnson2018b,Johnson2018c,Johnson2019,HeatEngines2019c,HeatEngines2019e}. 
  \par
  The objective of this paper is to define a new class of holographic heat engines using charged AdS black holes with logarithmic $U(1)$ gauge theory as a matter source. In fact, there are strong motivations for considering a logarithmic Lagrangian for the Abelian $U(1)$ gauge theory. In the logarithmic version of nonlinear electrodynamics (NED) theory (proposed by Soleng in \cite{NLBHSoleng1995}), the logarithmic action leads to a bounded field strength and so the electromagnetic self mass of a point charge will be finite \cite{NLBHSoleng1995}. Heisenberg and Euler have shown in their theory that, even in the vacuum, the Maxwell equations have to be exchanged by more fundamental theories of nonlinear electrodynamics in order to explain the vacuum polarization effects \cite{HeisenbergEuler1936}. Nonlinear electrodynamics Lagrangians such as Heisenberg-Euler theory, Born-Infeld (BI) theory \cite{BornInfeld} and logarithmic NED theory can be regarded as an effective field theories for simulating the 1-loop corrections of vacuum polarization (Feynman) diagrams computed in QED \cite{SchwartzQFT}. On the other hand, the quantum mechanical non-linearity of electromagnetic fields can be described by BI-type NEDs, which leads to scattering of light-by-light (Halpern scattering) \cite{JacksonElectrodynamics}. Note that the logarithmic NED is indeed a BI-type theory which can be manifest by expanding its Lagrangian in the weak-field coupling limit. In principle, it covers all the features of BI electrodynamics, so it is a viable theory which certainly merits further exploration. Moreover, in superstring theory, the effective actions describing nonlinear BI-type electrodynamics have been found for dynamics of D-branes \footnote{We also should emphasize that the dynamics of electromagnetic fields on the world-volumes of D-branes is exactly governed by the standard BI theory. Assuming that the nonlinear parameter is large enough, the other kinds of BI-type theories can naturally describe the same dynamics.} \cite{StringBI1,StringBI2,StringBI3,StringBI4}. For these reasons, corrections to black hole physics from nonlinear theories of electrodynamics as the matter field have been a subject of long-standing investigation \cite{NLBHs1935,NLBHs1937,NLBHSoleng1995,NLBHs1,NLBHs2,NLBHs3,NLBHs4,NLBHs5,NLBHs6,NLBHs7,NLBHs8,NLBHs9,NLBHs10,NLBHs11,NLBHs12,NLBHs13}. 
 \par

Logarithmic $U$(1) gauge theory of electrodynamics qualitatively shares many similarities with BI and Euler-Heisenberg theories in the weak-field limit, including describing 1-loop correction of the vacuum polarization, finite self-energy for point charges etc \cite{NLBHSoleng1995,BornInfeld,HeisenbergEuler1936,Berestetskii1,Berestetskii2,SchwartzQFT,LogEPJC}. Logarithmic, BI and Euler-Heisenberg gauge theories of electrodynamics have the following expansion in the weak-field limit (large enough $\cal \beta$)
 \begin{equation}
{\cal L}({\cal F}) =  - {\cal F} + {a_1}\frac{{{{\cal F}^2}}}{{{\beta ^2}}} - {a_2}\frac{{{{\cal F}^3}}}{{{\beta ^4}}} + O(1/{\beta ^6})
 \end{equation}
 where $a_i$'s are some positive constants (for example, see the expansion of the logarithmic $U(1)$ gauge theory in Sect. \ref{black holes}, Eq. (\ref{logarithmics})). So, it seems sensible that these theories share some common features (especially in the weak field limit), leading to what might be called ‘BI-type nonlinear electrodynamics’. However, BI-type gauge theories can behave differently in strong-field limit and the outcomes in this limit should be examined for each theory separately. Despite a series of similarities between these theories there are some differences which depend on the functional form of the Lagrangian density. Focusing on the cases of BI and logarithmic NED theories, the vacuum birefringence phenomenon is absent in BI theory while it is present in the logarithmic theory of electrodynamics \cite{LogEPJC}. As an another example, the finite self-mass/energy of a point charge in logarithmic and Born-Infeld theories are not the same (The latter is greater but the the maximal electrostatic field of the point charge in BI theory is smaller than the logarithmic electrodynamics \cite{BornInfeld,LogEPJC}). More interestingly, the study of full vacuum polarization diagrams in QED reveals that there is a logarithmic dependence on the cut-off in the effective action \cite{SchwartzQFT}. This is equivalent to the interesting finding of Euler and Heisenberg in which a logarithmic term of the field strength appears in the action as an exact 1-loop correction to the vacuum polarization \cite{HeisenbergEuler1936}.  Furthermore, it is shown in \cite{Berestetskii1,Berestetskii2} that for slowly varying fields (fields which cannot in practice create real electron-positron pairs, i.e., purely electromagnetic fields), the effective Lagrangian for radiative corrections increases at high field intensities logarithmically. Consequently, this observation shows that logarithmic electrodynamics is notable to examine photonic processes in certain regions of the electromagnetic fields. In addition, as indicated in \cite{NLBHSoleng1995}, the logarithmic U(1) gauge theory, which is contained in the class of theories constructed in \cite{Altshuler}, can be regarded as a possible mechanism for inflation. For these reasons, logarithmic U(1) gauge theory is not just a toy model with mathematical purposes and it has some physical evidences.

  Having these motivations, it would be interesting to examine the idea of holographic heat engines coupled with the logarithmic $U(1)$ gauge theory and investigate the effect of nonlinear matter source on them. However, in Ref. \cite{Johnson2016a}, spherically symmetric Born-Infeld AdS black holes as heat engines have been studied using high temperature expansion, but here, logarithmic $U(1)$ AdS black holes with various (spherical, planar and hyperbolic) symmetries are considered as a new class of holographic heat engines and our considerations are not restricted to the high temperature limit. We extensively explore the whole region of parameter space in low and high temperature limits for both the weak- and strong-field regimes. Elementally, in the weak coupling limit, our results approximately approach those of charged black hole heat engines with any kind of BI-type theories as matter source. In the strong coupling limit for any kind of BI-type theories, a qualitative behaviour similar to our results in this paper is expected. So, the results of this paper is predicted for all the BI-type theories. For our purposes, this paper is organized as follows: In Sec. \ref{black holes}, we define our set up and notation and give a brief review of Logarithmic $U(1)$ AdS black holes with their features. In Sec. \ref{Thermodynamics}, we give a brief review of the conserved charges, the extended black hole thermodynamics and the appearance of vacuum polarization quantity in a natural way. In Sec. \ref{Heat Engines}, we explore the idea of holographic heat engines, present the exact efficiency formula and investigate the engine's performance in weak and strong coupling regimes, followed by the numerical results and interesting figures. Finally in Sec. \ref{Final}, the results are summarized and discussed.

\section{Logarithmic $U(1)$ AdS black holes} \label{black holes}

The action of Einstein gravity coupled with the logarithmic $U(1)$ gauge field in AdS background is given by
 \begin{equation}
 {\cal{I}} =\frac{-1}{16 \pi}\int_{\cal{M}} d^D x\sqrt{-g}\big[R-2\Lambda+\mathcal{L(F)}\big],
 \end{equation}
 where $\Lambda=-(D-1)(D-2)/2L^2$ is the cosmological constant with  the AdS radius $L$. In the above action, $\mathcal{L(F)}$ is the Lagrangian of logarithmic form of BI-type theory \cite{NLBHSoleng1995}
 \begin{equation}\label{logarithmicsexact}
\mathcal{L(F)}= -8 \beta^2 \rm{ln} \Big(1+\frac{\cal{F}}{8\beta^2}\Big),
 \end{equation}
 where $\beta$ as a real constant denotes the non-linearity parameter and corresponds to the strength of the maximal electric field, ${\cal{F}}=F^{\mu \nu}F_{\mu \nu}$ is the Maxwell invariant, and $F^{\mu \nu}$ is the electromagnetic field strength tensor defined by $F_{\mu\nu}=\partial_\mu A_\nu-\partial_\nu A_\mu$. The non-linearity parameter, $\beta$, has units of electric intensity, $\rm{V/m}$, in SI units (in natural units with dimension of $mass$). In the weak coupling limit $\beta \to \infty$ (not infinity but large enough), ${\cal L}(\cal F)$ is of the form
 \begin{equation} \label{logarithmics}
{\left. {{\cal L}({\cal F})} \right|_{\beta  \to \infty }} =  - {\cal F} + \frac{{{{\cal F}^2}}}{{16{\beta ^2}}} - \frac{{{{\cal F}^3}}}{{192{\beta ^4}}} + O(1/{\beta ^6}).
 \end{equation}
The leading term of the expansion is the standard Maxwell's Lagrangian. The next leading term, $O({{\cal F}^2}/{\beta ^2})$, corresponds to the lowest order quantum correction to the classical Maxwell's Lagrangian. Therefore, the nonlinear parameter $\beta$ allows us to control the strength of the higher derivative corrections to the Abelian Maxwell field. However, in the strong coupling limit, i.e. $\beta \to 0$ (not zero but small enough), we are not allowed to use this expansion and the exact form of the logarithmic Lagrangian (\ref{logarithmicsexact}) should be applied.
 
The gravitational and electromagnetic field equations of Einstein gravity in the presence of logarithmic NED reads
\begin{equation} \label{Einstein field equations}
G_{\mu \nu}+\Lambda g_{\mu\nu}=\frac{1}{2}g_{\mu\nu}\mathcal{L(F)}-2F_{\mu\lambda}F_{\nu}\,^{\lambda}\frac{\partial\mathcal{L(F)}}{\partial\mathcal{F}},
\end{equation}
and
\begin{equation}\label{EMFE}
\partial_{\mu}\Big(\sqrt{-g}\frac{\partial\mathcal{L(F)}}{\partial\mathcal{F}} F^{\mu\nu}\Big)=0.
\end{equation}
For the spacetime metric, the following static ansatz with spherical symmetry for the event horizon is used
\begin{equation}
d{s^2} =  - f(r)d{t^2} + \frac{{d{r^2}}}{{f(r)}} + {r^2}\big( {d x _1^2 + \sum\limits_{i = 2}^{D - 2} {\prod\limits_{j = 1}^{i - 1} {{{\sin }^2}{x _j}d x _i^2} } } \big).
\end{equation}
In what follows, we use the symbol "${\Sigma _{D-2}}$" as the volume of the unit $(D-2)$-sphere, given by \footnote{For hyperbolic and planar black holes, thermodynamic quantities can be computed per volume, $\Sigma _{D - 2}$. However, as will be seen in Sect. \ref{Heat Engines}, the volume $\Sigma _{D - 2}$ will not appear in the efficiency formula.}
\begin{equation}
{\Sigma _{D - 2}} = \frac{{2{\pi ^{(D- 1)/2}}}}{{\Gamma \left( {\frac{{D - 1}}{2}} \right)}},
\end{equation}
where ${\Gamma \left( {\frac{{D - 1}}{2}} \right)}$ is the gamma function. We also generalize our considerations to AdS black holes with a horizon of zero (denoted by $k=0$) and negative (denoted by $k=-1$) constant curvatures. At the end, the cases with planar and hyperbolic black holes as heat engines will be discussed. The gauge field $A_\mu$ is determined by solving the electromagnetic field equation \ref{EMFE}. Assuming the electrostatic ansatz as ${A_\mu } = \Phi (r)\delta _\mu ^t$, the differential equation of scalar potential ($\Phi$) is obtained as follows
\begin{equation} \label{gauge potential}
\frac{r}{{{\beta ^2}}}\,\partial _r^2\Phi (r)\,{\left( {{\partial _r}\Phi (r)} \right)^2} - \frac{{(D - 2)}}{{{\beta ^2}}}{\left( {{\partial _r}\Phi (r)} \right)^3} + 4r\,\partial _r^2\Phi (r) + 4(D - 2){\partial _r}\Phi (r) = 0.
\end{equation}
The first two terms stem from the logarithmic $U(1)$ electrodynamics. As $\beta \to \infty$, these terms approaches zero and the differential equation reduces to the Maxwell case. Solving equation \ref{gauge potential} yields
\begin{equation}\label{gauge potential_ans}
\Phi (r) = \frac{{2{\beta ^2}{r^{D - 1}}}}{{(D - 1)q}}\left(1- {}_2{F_1}\left( {\left[ { - \frac{1}{2},\,\frac{{ - (D - 1)}}{{2(D - 2)}}} \right],\,\left[ {\frac{{D - 3}}{{2(D - 2)}}} \right],\,\Upsilon (r) } \right) \right)+C_1
\end{equation}
where 
\begin{equation} \label{Upsilon}
\Upsilon (r)  =  - \dfrac{q^2 }{\beta ^2r^{2(D - 2)}},
\end{equation}
and $C_1$ is a constant which is specified by a gauge fixing (see the next section). Then, having the gauge vector field $A_\mu$, we can easily obtain the field strength tensor, $F_{\mu \nu}$. The only non-zero components of the field strength tensor are $F_{rt}=-F_{tr}$. This determines the associated electrostatic field as $F_{rt} = {\nabla _r}\Phi (r)=E(r)$. Accordingly, in this gauge theory, the electrostatic field of a point particle (or outside the event horizon of a charged black hole) is obtained as
\begin{equation} \label{electric field}
E(r) = \frac{2}{{1 + \sqrt {1 + \frac{{{q^2}}}{{{\beta ^2}{r^{2(D - 2)}}}}} }}\frac{q}{{{r^{D - 2}}}}.
\end{equation}
In remote distances ($r \to \infty$), the standard Coulomb behaviour, $E=q/r^{(D-2)}$, is recovered, no matter how much is the strength of non-linearity coupling $\beta$ (the power of $r_+$ is greater than the power of $\beta$).

Having the stress-energy tensor, we can construct the equation of motion for the gravity side, equation (\ref{Einstein field equations}), which admits the emblackening factor, $f(r)$, given by
\begin{eqnarray}\label{fr}
f(r)=&& k - \frac{m}{{{r^{D - 3}}}} + \frac{{{r^2}}}{{{L^2}}} + \frac{{8{r^2}\beta }}{{\left( {D - 2} \right){{\left( {D - 1} \right)}^2}}}\left( {\left( { - 2D + 3} \right){r^{2 - D}}\xi (r) + \beta \left( {2D - 3 - \left( {D - 1} \right)\ln \left[ {\frac{{2{r^D}\beta }}{{{r^D}\beta  + {r^2}\xi (r)}}} \right]} \right)} \right)\nonumber\\
 &&+ \frac{{8\left( {D - 2} \right){q^2}{r^{8 - d}}\beta \sqrt {1 - \Upsilon }\, \xi (r)}}{{\left( {D - 3} \right){{\left( {D - 1} \right)}^2}\left( {{q^2}{r^4} + {r^{2D}}{\beta ^2}} \right)}}{}_2{F_1}\left( {\left[ {\frac{1}{2},\frac{{D - 3}}{{2\left( {D - 2} \right)}}} \right],\left[ {\frac{{ 3D-7}}{{2D-4}}} \right],\Upsilon (r) } \right)
\end{eqnarray}
with
\begin{equation} \label{Xi}
\xi (r)=\sqrt{q^2 + r^{2(D-2)} \beta^2}.
\end{equation}
It is assumed that the black hole's horizon could have spherical ($k=+1$), planar ($k=0$) or hyperbolic ($k=-1$) symmetries in obtaining the emblackening factor $f(r)$. Outstandingly, checking the Kretschmann invariant ($K = {R_{abcd}}{R^{abcd}}$) shows that the singularity at the origin ($r=0$) is much weaker than the singularities of the conventional linear Maxwell cases \cite{NLBHSoleng1995,NLBHs4}. In four spacetime dimensions, the Kretschmann scalar behaves as $K \propto 1/{r^4}$ for the logarithmic $U(1)$ case, while in the linear Maxwell limit ($\beta=\infty$) it behaves as $K \propto 1/{r^8}$. Moreover, the black hole solutions possesses an event horizon ($r_+$) which is the largest root of $f(r_+)=0$.

\section{Conserved charges, thermodynamics and vacuum polarization} \label{Thermodynamics}
In this section, we briefly review the conserved quantities of the logarithmic $U(1)$ AdS black hole solutions with different topologies for the event horizon and then verify the first law of thermodynamics and the associated Smarr relation in the extended phase space. Let us start with the finite mass. In asymptotically AdS backgrounds, one can use the Komar mass integral or the conformal method of Ashtekar-Magnon-Das (ADM) to obtain the well-known ADM mass formula as \cite{Ashtekar1984,Ashtekar2000}
\begin{equation}
M = \frac{{{\Sigma _{D - 2}}(D - 2)}}{{16\pi }}m,
\end{equation}
in which the constant $m$ is a parameter related to the total mass and it is obtained from $f(r_+)=0$. Hence the mass or more accurately the enthalpy, is given by
\begin{eqnarray}\label{MADM}
M =&& \frac{{\left( {D - 2} \right){\Sigma _{D - 2}}}}{{16\pi }}\Bigg( 
k {r_ + ^{D - 3}} + \frac{{2 \Lambda {r_ + ^{D - 1}}}}{{{(D-1)(D-2)}}} - \frac{{8\beta \left( {\left( {2D - 3} \right){r_ + }\xi_+  + {r_ + ^{D-1}}\beta \left( {3 - 2D +\ln (2) \left( {D - 1} \right)} \right)} \right)}}{{{{\left( {D - 1} \right)}^2}(D - 2)}}\nonumber\\
 &&- \frac{{8{r_ + ^{D - 1}}{\beta ^2}\ln \Big( { - \frac{{{r_ + ^{D - 4}}\beta \left( {{r_ + ^D}\beta  - {r_ + ^2}\xi_+ } \right)}}{{{q^2}}}} \Big)}}{{(D - 1)(D - 2)}} + \frac{{8\left( {D - 2} \right){q^2}{r_ + ^{3 - D}} {}_2{F_1}\left( {\left[ {\frac{1}{2},\frac{{D - 3}}{{2(D - 2)}}} \right],\left[ {\frac{{3D - 7}}{{2D - 4}}} \right],\Upsilon_+ } \right)}}{{\left( {D - 3} \right){{\left( {D - 1} \right)}^2}}}
 \Bigg),
\end{eqnarray}
where the conventions $\xi_+=\xi(r_+)$ and $\Upsilon_+=\Upsilon (r_+)$ have been used for relations (\ref{Xi}) and (\ref{Upsilon}). The Hawking temperature of the logarithmic $U(1)$ AdS black holes can be obtained by use of the definition of surface gravity \cite{Hawking1975}, ${\kappa ^2} = -\frac{{ 1}}{2}({\nabla _\mu }{\chi _\nu })({\nabla ^\mu }{\chi ^\nu })$ where ${\chi _\mu } = \delta _\mu ^t{\partial _t}$ is the timelike Killing field, yielding 
\begin{eqnarray}\label{temp}
T &=& \frac{\kappa}{2 \pi}= \frac{1}{{4\pi }}{\left. {\frac{{\partial f(r)}}{{\partial r}}} \right|_{r = {r_ + }}} \nonumber \\
&=&\frac{{(D - 2)(D - 3)k - 2\Lambda r_ + ^2 + 8{\beta ^2}r_ + ^2\left( {\ln \Big( {\frac{{1 + \sqrt {1 + \Upsilon_+ } }}{2}} \Big) + 1 - \sqrt {1 + \Upsilon_+ } } \right)}}{{4\pi (D - 2){r_ + }}}.
\end{eqnarray}
The entropy is given in terms of the horizon radius via the well-known Bekenstein-Hawking formula as
\begin{equation} \label{entropy}
S =\dfrac{A}{4}= \frac{{{\Sigma _{D - 2}}}}{4}r_ + ^{D - 2}.
\end{equation}

In order to compute the $U(1)$ charge, one can calculate the charge passing through a $n$-dimensional hypersphere at spatial infinity with the same geometry as the event horizon \cite{Carroll2004}. Therefore, the total charge is given by the generalized Gauss' law as
\begin{equation} \label{charge}
Q =  - \frac{1}{{4\pi }}\int_{{r_\infty }} {{d^{D-2}}x \left( {\frac{{\partial {\cal L}({\cal F})}}{{\partial {\cal F}}}} \right){n_\mu }{\sigma _\nu }{F^{\mu \nu }}}  = \frac{{{\Sigma _{D - 2}}}}{{4\pi }}q,
\end{equation}
in which $n_\mu$ and $\sigma_\mu$ are (outward-pointing) unit normal vectors defined as
\begin{equation}
{n_\mu } =  - \sqrt {f(r)} dt\,,\,\,\,\,\,\,{\sigma _\mu } = \frac{1}{{\sqrt {f(r)} }}dr.
\end{equation}
The $U(1)$ potential conjugate to the electric charge can be obtained using the elementary discussions in classical electrodynamics \cite{JacksonElectrodynamics}.  We work in a gauge in which the norm of the gauge field, $A^2={A_\mu} {A^\mu}$, is finite at the horizon. This leads to ${A_t}(r_+)=0$ at the horizon and the constant $C_1$ in equation (\ref{gauge potential_ans}) is obtained as 
\begin{equation}
C_1=- \frac{{2{\beta ^2}{r_+^{D - 1}}}}{{(D - 1)q}}\left(1- {}_2{F_1}\left( {\left[ { - \frac{1}{2},\,\frac{{ - (D - 1)}}{{2(D - 2)}}} \right],\,\left[ {\frac{{D - 3}}{{2(D - 2)}}} \right],\,\Upsilon_+ } \right) \right).
\end{equation}
Hence, using the fact that $E(r) = {\nabla _r}\Phi (r)$, the gauge potential with respect to the event horizon is calculated as 
\begin{eqnarray}
\Phi  = \Phi (\infty ) - \Phi ({r_ + }) &=& \int_{{r_ + }}^\infty  {E(r) dr} \nonumber \\
&=&  - \frac{{2{\beta ^2}r_ + ^{D - 1}}}{{(D - 1)q}}\left( {1 - {}_2{F_1}\left( {\left[ { - \frac{1}{2},\,\frac{{ - (D - 1)}}{{2(D - 2)}}} \right],\,\left[ {\frac{{D - 3}}{{2(D - 2)}}} \right],\,\Upsilon_+ } \right)} \right).
\end{eqnarray}

It is time to invoke the main idea of the extended phase space, interpreting the mass as enthalpy ($H \equiv M$) by treating $\Lambda$ as pressure ($\Lambda =- 8 \pi P$). With this picture in mind, the thermodynamic volume conjugate to $ P $ is naturally obtained as
\begin{equation}\label{vol}
V = {\left( {\frac{{\partial H}}{{\partial P}}} \right)_{S,Q}} = \frac{{{\Sigma _{D - 2}}}}{{D - 1}}r_ + ^{D - 1}.
\end{equation}
It is not difficult to show that the temperature and the potential can also be computed by using the following thermodynamic relations
\begin{equation}
T = {\left( {\frac{{\partial H}}{{\partial S}}} \right)_{P,Q}}\,,\,\,\,\,\Phi  = {\left( {\frac{{\partial H}}{{\partial Q}}} \right)_{S,P}},
\end{equation}
that means these quantities satisfy the first law of thermodynamics in the extended phase space as
\begin{equation}\label{Efirstlaw}
dH = TdS + VdP + \Phi dQ.
\end{equation}
Moreover, these quantities satisfy the following Euler-type relation 
\begin{equation}
(D-3)H=(D-2)TS-PV+(D-3) \Phi Q - {\mathcal B} \beta
\end{equation}
which is called Smarr formula in community. The new thermodynamic quantity $\cal B$ is conjugate to the non-linearity parameter $\beta$ and is of the form
\begin{eqnarray} \label{vacuum polarization}
\mathcal{B} =&& {\left( {\frac{{\partial H}}{{\partial \beta}}} \right)_{S,P,Q}}\nonumber\\
=&&
\frac{{{\Sigma _{D - 2}}}}{{2\pi {{\left( {D - 1} \right)}^2}\xi_+ }}\Bigg( 
\left( {5 - 3D} \right){\beta ^2}{r_{+}^{2D - 3}} + \left( {5 - 3D} \right){q^2}r_{+} - \beta  \left( {5 - 3D + \ln \left( 4 \right)\left( {D - 1} \right)} \right) \xi_+ {r_{+}^{D - 1}} \nonumber\\
&&+ \frac{{\left( {D - 2} \right){q^2}{r_{+}^{1 - 2D}}\left( {{q^2}{r_{+}^4} + {\beta ^2} {r_{+}^{2D}}} \right){}_2{F_1}\left( {\left[ {\frac{1}{2},\frac{{D - 3}}{{2\left( { D-2} \right)}}} \right],\left[ {\frac{{3D - 7}}{{2D - 4}}} \right],\Upsilon_+ } \right)}}{{{\beta ^2}\sqrt {1 - \Upsilon_+ } }}\nonumber\\
&& - 2 (D - 1) \beta {r_{+}^{D - 1}}\xi_+ \ln \Big( {\frac{{\beta{r_{+}^{D - 4}} \left( {{r_{+}^2}\xi_+ -\beta {r_{+}^D} } \right)}}{{{q^2}}}} \Big)\Bigg).
\end{eqnarray}

\begin{figure}[!htbp]
	\epsfxsize=9 cm 
	\includegraphics[width=9 cm]{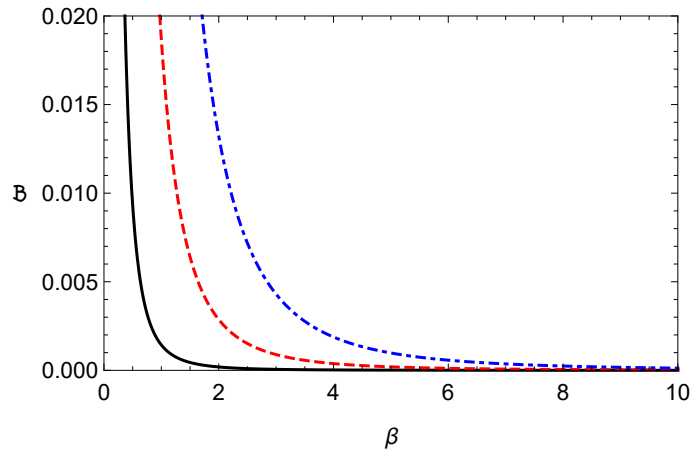}
	\caption{The logarithmic vacuum polarization, ${\cal B}$, versus $\beta$  for $r_{+}=1$, $D=4$,  and $q=0.5$ (black solid line), $q=1$ (red dashed line) and $q=1.5$ (blue dot-dashed line).}
	\label{VacPol_beta}
\end{figure}

\begin{figure}[!htbp]
	\epsfxsize=8.5 cm
	\includegraphics[width=9 cm]{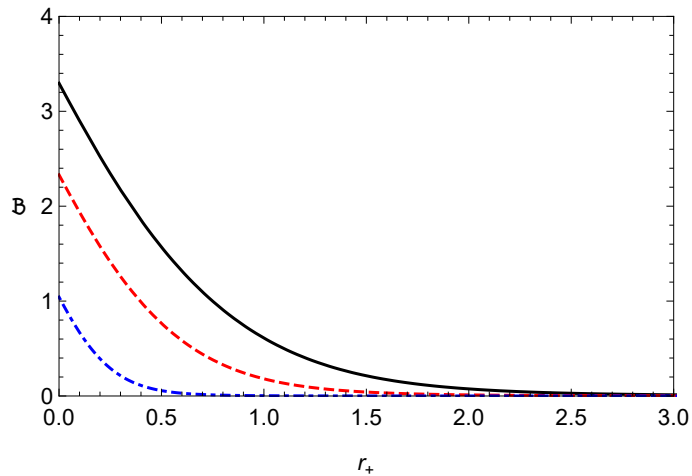}
	\caption{The logarithmic vacuum polarization, ${\cal B}$, versus $r_+$ for $D=4$, $q=2$ and $\beta=0.5$ (black solid line),  $\beta=1$ (red dashed line) and  $\beta=5$ (blue dot-dashed line).}
	\label{VacPol_r}
\end{figure}

Obviously, the non-linearity parameter $\beta$ is necessary to satisfy the Smarr relation. On the other hand, it is required for consistency of both the extended first law and the corresponding Smarr relation. So we should consider $\beta$ as a thermodynamic phase space variable in what follows, yielding
\begin{equation}\label{EfirstlawwithBeta}
dH = TdS + VdP + \Phi dQ + {\cal B} d\beta.
\end{equation}
Therefore, considering the logarithmic NED in the extended phase space leads to the appearance of  a new thermodynamic quantity ($\cal B$) which has a thermodynamic interpretation. This quantity also appears in the Smarr formula of AdS black holes with a BI NED source and it was referred to as the BI \textit{vacuum polarization} in Ref. \cite{Mann2012}. It is inferred that the appearance of this quantity ($\cal B$) is the characteristic of any BI-type nonlinear matter source which is absent in other nonlinear $U(1)$ gauge field theories such as power Maxwell invariant theories of Electrodynamics. Therefore, following \cite{Mann2012}, $\cal B$ is called logarithmic vacuum polarization by us. It is instructive to seek about the physical meaning of this quantity. As indicated in \cite{Mann2012}, ${\cal B}$ has units of electric polarization since the term ${\cal B}\beta$ has units of energy and $\beta$ has units of electric field strength. The limit $\beta = \infty$ corresponds to Maxwell Electrodynamics, so the vacuum polarization vanishes in this limit. This is one of the main differences between Maxwell Electrodynamics and any BI-type NED theory. In the weak-field limit ($\beta \to \infty$), $\cal B$ approaches zero rapidly by further increasing $\beta$ (this happens for BI theory as well \cite{Mann2012}). However, as depicted in Fig. \ref{VacPol_beta}, it becomes larger in the strong coupling regime and, eventually, diverges at $\beta = 0$. As will be shown in the next section, the weak-field limit is mathematically equal to long distances in which the event horizon radius is large and so we are far enough from the nonlinear matter source for applying the expansion. This can simply be inferred from Figs. \ref{VacPol_beta} and \ref{VacPol_r}, in which $\cal B$ approaches zero for large enough values of $r_+$ or $\beta$, indicating that the expansion of all quantities in this paper around large $r_+$ is the same as expansion around large $\beta$. In addition, for both the strong and weak coupling regimes, logarithmic vacuum polarization for a fixed $\beta$ reaches a finite value at the origin. This behaviour is illustrated in Fig. \ref{VacPol_r}. \\

\section{Holographic Heat engines} \label{Heat Engines}
\begin{figure}[!htbp]
	\epsfxsize=5 cm 
\includegraphics[width=6 cm]{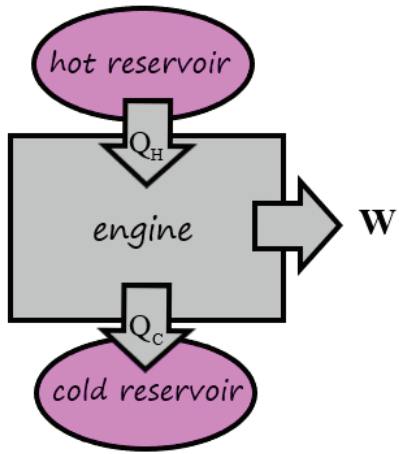}
	\caption{Energy-flow diagram for a heat engine.}
	\label{he}
\end{figure}

\begin{figure}[!htbp]
	\epsfxsize=9 cm 
	\includegraphics[width=9 cm]{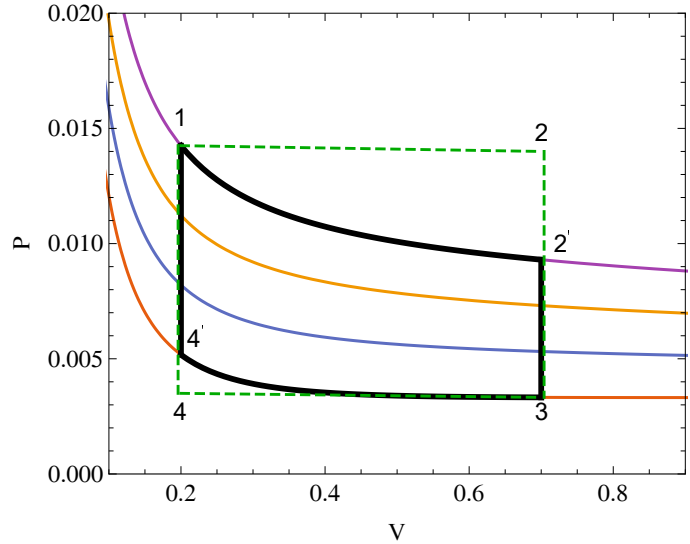}
	\caption{The black solid loop (points $1-2^{'}-3-4^{'}$) shows the Carnot cycle which is the same as Stirling cycle for static black holes. The dashed green rectangular cycle (points $1-2-3-4$) is an ideal engine for which the efficiency is computed. }
	\label{carnot}
\end{figure}

Having the extended first law of thermodynamics (\ref{EfirstlawwithBeta}) in hand, the black hole can be treated as a working substance of a classical heat engine, that absorbs heat $Q_H$, produces mechanical work $W$ and dumps waste heat $Q_C$ into the environment (See Fig. \ref{he}). The efficiency of an engine which is defined as $\eta=W/Q_H=1-Q_C/Q_H$ depends on the choice of the paths of the engine cycle in the $P-V$ plane and also the equation of state of the black hole considered as a working substance. Imposing the second law of thermodynamics, the maximum efficiency that can be achieved by an engine is $\eta_C=1-T_C/T_H$. This is the efficiency of a Carnot cycle consisting two adiabats and two isotherms. Labeling the isotherm with higher temperature as $T_H$ and the one with lower temperature by $T_C$, the Carnot cycle operates in four steps: First an isothermal expansion at $T_H$ through which $Q_H$ is absorbed, second an adiabatic expansion to $T_C$, third an isothermal compression at $T_C$ while dumping extra heat $Q_C$ and at last an adiabatic compression bringing back the substance to $T_H$ (See Fig. \ref{carnot}). During the Carnot cycle, no new entropy is created and therefore the efficiency always possesses the maximum value allowed by thermodynamics' laws,  no matter what is the working substance  or what is the equation of state that governs it. For this reason it is logical to consider the Carnot efficiency as a benchmark and compare efficiencies of our desired black hole heat engines with it. 

\par
{It can clearly be seen from Eqs. (\ref{entropy}) and (\ref{vol}) that both the entropy and the volume depend only on the horizon radius $r_{+}$ and indeed they are not independent quantities, meaning that for static black holes there is no difference between adiabats and isochores. As a consequence, Stirling and Carnot cycles are identical (See Fig. \ref{carnot}). Also for static black holes some well-known thermodynamic cycles such as Otto cycle (involving a sequence of adiabatic $\to$ isochoric $\to$ adiabatic  $\to$ isochoric processes) and Diesel cycle {(composed of adiabatic $\to$ isobaric $\to$ adiabatic $\to$ isochoric processes) have no place. Besides, for static black holes, the specific heat defines as
\begin{equation} \label{specific heat}
C=T{\left( {\frac{{\partial S}}{{\partial T}}} \right)} = T\left( {\frac{{\partial S}}{{\partial {r_ + }}}} \right)\left( {\frac{{\partial {r_ + }}}{{\partial T}}} \right),
\end{equation}
which vanishes at constant volume. For the case of logarithmic $U(1)$ AdS black holes, this can be checked by evaluating specific heat using Eqs. (\ref{temp}) and (\ref{entropy}) as
\begin{eqnarray}\label{specific}
C=&&{{\Sigma _{D - 2}}{r_ + ^{D - 2}}\left( {4{r_ + }\frac{{\partial P}}{{\partial T}} - D +2} \right)}\nonumber\\
&& \times \left( {\frac{{k(D-2)(D-3) - 8{\beta ^2}{r_ + ^2}\left( { - 1 + \sqrt {1 - \Upsilon_+ } } \right) + 8{\beta ^2} {r_ + ^2} \ln \left( {\frac{1}{2}\left( {1 + \sqrt {1 - \Upsilon_+ } } \right)} \right) + 16\pi  P {r_ + ^2}}}{{{{4 (D - 3)\left( {k (D - 2) - 8 {\beta ^2} {r_ + ^2} \left( { - 1 + \sqrt {1 - \Upsilon_+ } } \right)} \right)}} - 32{\beta ^2}{r_ + ^2}\ln \left( {\frac{1}{2}\left( {1 + \sqrt {1 - \Upsilon_+ } } \right)} \right) - 64\pi P {r_ + ^2}}}} \right),
\end{eqnarray}
and then substituting ${\partial S}/{\partial r_+} = 0$ into Eq. (\ref{specific heat}) or ${\partial P}/{\partial T}=(D-2)/4r_{+}$ into Eq. (\ref{specific}) which leads to $C_V=0$. The specific heat at constant pressure $C_P$ is simply given by setting ${\partial P}/{\partial T}=0$ in Eq. (\ref{specific}). Now, having the explicit form of $C_P$ with suitable choice of the desired cycle, it would be simpler to study  the engine's efficiency. Fig. \ref{carnot} shows the rectangular cyclic process constructed out of a pair of isobaric and a pair of isochoric processes, often called the ideal cycle.  If the black hole undergoes the ideal cycle, the absolute work done can be written as
\begin{equation}\label{work}
W= W_{1\rightarrow 2}+W_{3\rightarrow 4}=(P_1-P_4)(V_2-V_1),
\end{equation}
 and the heat flows into the loop through the top isobar is given by
 \begin{equation}\label{heat}
 Q_H=\int_{T_1}^{T_2}C_P(P_1,T)dT.
 \end{equation}
 Solving for $r_{+}$ in terms of $T$ using Eq. (\ref{temp}) and substituting in Eq. (\ref{specific}), $C_P$ can be rewritten as a function of $T$ and the integration along the isobar in order to calculate $Q_H$ can be performed. For most cases, this integration is complicated and, as a result, a high temperature limit can be studied by performing a series expansion for $T$ in Eq. (\ref{temp}) about large $r_{+}$. This was done before for Born-Infeld \cite{Johnson2016a} and Gauss-Bonnet \cite{Johnson2016b} AdS black holes and the engine efficiencies were found in the high temperature limit. We will also study the high temperature limit for the logarithmic charged AdS black hole in detail in section \ref{secht}.
 \par  
 For now, we use a much more straightforward way to evaluate efficiency which was first proposed in \cite{Johnson_exact}. The first law of thermodynamics for black holes (\ref{Efirstlaw}) suggests a simple way to compute heat flows ($Q_H$ and $Q_C$) along the isobars in the ideal cycle of Fig. \ref{carnot}. For the isobaric curves, the pressure is constant ($dP=0$) and, as a result, the thermodynamic identity leads to $dH=TdS$ for a black hole engine with specified charge as well as nonlinear parameter ($\beta$), which means that during these constant-pressure processes, the heat flows cause the enthalpy to change and the compression-expansion work has no effect. In other words, having the enthalpy changes, the heat flows and consequently the efficiency of the black hole engine are at hand. Since the enthalpy of the AdS black hole equals its mass in the extended phase space, thus the efficiency is obtained in terms of the black hole mass as
 \begin{equation}\label{EEF}
 \eta=1-\frac{M_3-M_4}{M_2-M_1},
 \end{equation} 
 where $1,2,3$ and $4$ refer to the mass of black hole evaluated at the corners of the rectangular cycle in Fig. \ref{carnot}.
 \par
 Before proceeding further, it should be noted that the equation of state of logarithmic $U(1)$ AdS black hole, i.e.,
\begin{eqnarray}
P &=& \frac{{(D - 2)T}}{{4{r_ + }}} - \frac{{(D - 2)(D - 3)k}}{{16\pi r_ + ^2}} - \frac{{{\beta ^2}}}{{2\pi }}\left( {\ln \left( {\frac{{1 + \sqrt {1 + \Upsilon_+} }}{2}} \right) + 1 - \sqrt {1 + \Upsilon_+} } \right);\nonumber\\
{r_ + } &=&  {\left( {\frac{{(D - 1)V}}{{{\Sigma _{D - 2}}}}} \right)^{\frac{1}{{D - 1}}}},
\end{eqnarray}
shows the possibility of phase transitions for sufficiently low temperatures. As an example, a  characteristic feature of isotherms in the $P-V$ plane for the well-known Van der Waals behaviour is displayed in Fig. \ref{eos}. No phase transition is seen for high temperature isotherms ($T>T_{cr}$) and so the phase boundary in the $P-T$ plane vanishes above the critical temperature, $T_{cr}$. For $T<T_{cr}$, as in the Van der Waals liquid-gas system, a first order phase transition occurs, here, between small and large black holes. Note that in this region the corrections should be implemented to the oscillatory parts of isotherms, in very much the same way to Maxwell construction for the Van der Waals fluid and, as a result, by decreasing pressure the system will go straight from the small black hole state to the large state and vice versa with an abrupt decrease/increase in the volume. Therefore the unphysical pressure regions and also the multi-valued parts disappear. Working in the safe temperature domain, i.e. high enough to avoid multivaluedness corresponding to phase transitions, allows us to consider heat engine cycles in $T>T_{cr}$ region. Hence we do not focus on phase transitions here (See Refs. \cite{Kubiznak2012,Mann2012,vdWNED2019} for more details about the subject of black hole phase transitions). It is worthwhile to point that the leading order of the equation of state in the high temperature limit reads
\begin{equation}\label{eos_HT}
PV^{\frac{1}{D-1}}\sim T,
\end{equation}
which is the ideal gas limit for our black holes. However, in what follows, we will study the whole parameter space of the theory including low and high temperatures. In the low temperature domain, the parameters of our black hole engines have been picked up in such a way that no phase transition takes place at all.

\begin{figure}[!htbp]
	\epsfxsize=9 cm
\includegraphics[width=9 cm]{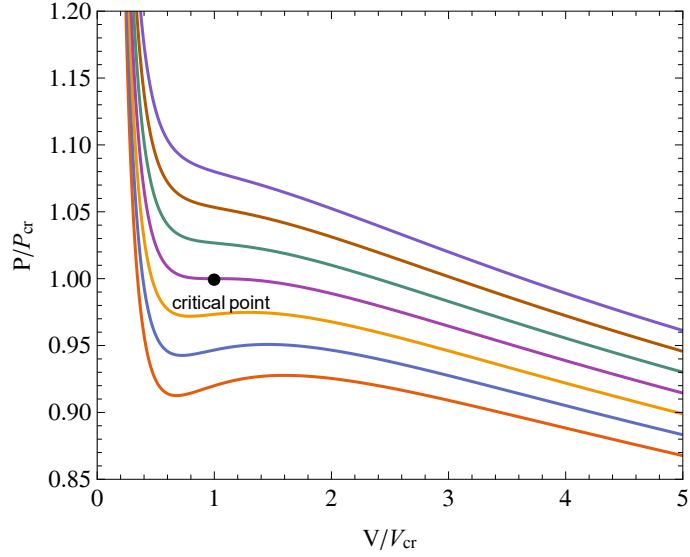}
	\caption{Isotherms in the $P-V$ plane for $D=4$, $q=1$, $\beta=10$. From bottom to top the isotherms are for $T/T_c$ ranging from $0.97$ to $1.03$ in increments of $0.01$. }
	\label{eos}
\end{figure}
\subsection{The exact efficiency formula for the heat engines: general considerations}
In this section we use the exact efficiency formula \ref{EEF} to verify the efficiency of the ideal cycle of Fig. \ref{carnot} for logarithmic $U(1)$ AdS black holes. Inserting Eq. (\ref{MADM}) into Eq. (\ref{EEF}), we obtain
\begin{eqnarray}\label{exactf}
\eta=\frac{N(r_1,r_2,P_1,P_3)}{D(r_1,r_2,P_1,P_3)},
\end{eqnarray}
  where
 \addtocounter{equation}{-1}
 \begin{subequations}
 \begin{align}
  N(r_1,r_2,P_1,P_3)=& 16\pi \beta \left( {D - 3} \right)\left( {D - 1} \right)\left( {{P_1} - {P_3}} \right){\left( {{r_1}{r_2}} \right)^{2(D + 1)}}\left( {{r_1}{r_2^D} - {r_1^D}{r_2}} \right),\\
  D(r_1,r_2,P_1,P_3)=&
   - 8{\left( {D - 2} \right)^2}{q^2}{r_1^8}{r_2^{2D + 3}}\sqrt {{q^2} + {r_1^{2D - 4}}{\beta ^2}} {}_2{F_1}\left( {{\Upsilon _1}} \right) \nonumber \\
  &+  {r_1^{2D}}\Bigg\{8{\left( {D - 2} \right)^2}{q^2}{r_1^3{r_2^8}}\sqrt {{q^2} + {r_2^{2D - 4}}{\beta ^2}} {}_2{F_1}\left( {{\Upsilon _2}} \right) \nonumber \\
   &+ \left( {D - 3} \right){r_2^{2D}}\beta \bigg(k\left( {D - 2} \right){\left( {D - 1} \right)^2}\left( { - {r_1^D{r_2^3}} + {r_1^3}{r_2^D}} \right) \nonumber \\
   &+ 8\left( {D - 1} \right){r_1^2}{r_2^2}{\beta ^2}\left( {{r_1^D}{r_2}\ln \left( {\frac{{2{r_1^D}\beta }}{{{r_1^D}\beta  + {r_2^2}{\xi _1}}}} \right) - {r_1}{r_2^D}\ln \left( {\frac{{2{r_2^D}\beta }}{{{r_2^D}\beta  + {r_2^D}{\xi _2}}}} \right)} \right)\nonumber \\
   & + 8{r_1^2}{r_2^2}\Big( {2\pi \left( {D - 1} \right){P_1}\left( { -{ r_1^D}{r_2} + {r_1}{r_2^D}} \right) + \left( {2D - 3} \right)\beta \left( { - {r_1^D}{r_2}\beta  + {r_1}{r_2^D}\beta  + {r_1^2}{r_2}{\xi _1} - {r_1}{r_2^2}{\xi _2}} \right)} \Big) 
   \bigg)\Bigg\}
 \end{align}
  \end{subequations}
with $\Upsilon_{1,2}=\Upsilon|_{r=r_1,r_2}$,  $\xi_{1,2}=\xi|_{r=r_1,r_2}$ and
 \begin{equation}
{}_2{F_1}({\Upsilon})={}_2{F_1}\left( {\left[ {1,\frac{{2D - 5}}{{2D - 4}}} \right],\left[\frac{3D-7} {2D - 4} \right],\Upsilon(r) } \right).\nonumber
\end{equation}

 As discussed earlier, it is reasonable to compare the efficiency of the ideal cycle with the efficiency of the Carnot engine, $\eta_{C}$. Also, it is worthwhile to compare the efficiency of logarithmic $U(1)$ AdS black hole heat engine with the efficiency of the Reissner-Nordstr\"{o}m AdS black hole which is the simplest holographic heat engine, since our black holes have deviations from Reissner-Nordstr\"{o}m AdS black holes in the electromagnetic field sector. We denote the efficiency in the Einstein-Maxwell limit with $\eta_0$ which is defined as $\eta_0=\lim_{\beta\rightarrow \infty } \eta(\beta)$. In order to obtain the efficiency in this limit, we should find the asymptotic behaviour of the finite mass, given by
  \begin{equation}\label{massinf}
  M|_{\rm{large}\,\beta}=\frac{{{\Sigma _{D - 2}}\left( {D - 2} \right)}}{{16\pi }}\left(k {{r_+^{D - 3}} + \frac{16 \pi P{{r_+^{D - 1}}}}{{{(D-1)(D-2)}}} + \frac{{2{q^2}{r_+^{3 - D}}}}{{\left( {D - 3} \right)\left( {D - 2} \right)}} - \frac{{{q^4}{r_+^{7 - 3D}}}}{{4\left( {D - 2} \right)\left( {3D - 7} \right){\beta ^2}}}} \right),
  \end{equation}
  which clearly goes to the Reissner-Nordstr\"{o}m ADM mass as $\beta$ goes to infinity. Inserting Eq. \ref{massinf} into the exact efficiency formula (\ref{EEF}) leads to
  \begin{equation}\label{effbetainf}
  \eta|_{\rm{large}\,\beta} =\frac{A(r_1,r_2,P_1,P_3)}{B(r_1,r_2,P_1,P_3)},
  \end{equation}
  where
\addtocounter{equation}{-1}
 \begin{subequations}
 \begin{align}
  A(r_1,r_2,P_1,P_3) =\,& 64\pi \left( {D - 3} \right)\left( {3D - 7} \right)\left( {{P_1} - {P_3}} \right)\left(- {r_1^D} r_2 + {r_1}{r_2^D} \right){\beta ^2},\\
B(r_1,r_2,P_1,P_3)=&
 \left( {D - 3} \right)\left( {D - 1} \right){q^4}{r_1}{r_2}\left( {r_1^{7 - 3D}} - {r_2^{7 - 3D}} \right) \nonumber\\
   & - 8\left( {D - 1} \right)\left( {3D - 7} \right){q^2}{r_1}{r_2}\left( {r_1^{3 - D}} - {r_2^{3 - D}} \right){\beta ^2}\nonumber\\
  &+ 64\pi \left( {D - 3} \right)\left( {3D - 7} \right){P_1}\left( { - {r_1^D}{r_2} + {r_1}{r_2^D}} \right){\beta ^2} \nonumber\\
  &+ 4k\left( {D - 3} \right)\left( {D - 2} \right)\left( {D - 1} \right)\left( {3D - 7} \right)\left( { - {r_1^{D - 2}}{r_2} + {r_1}{r_2^{D - 2}}} \right){\beta ^2}
  \end{align}
  \end{subequations}
   To check for correctness of this relation, one can easily take the limit $\beta \to \infty$ which yields the efficiency of the rectangular engine cycle for the $D$-dimensional Reissner-Nordstr\"{o}m AdS black hole as
  \begin{equation}
  \eta\,_{\rm{R.N. }}=\frac{C(r_1,r_2,P_1,P_3)}{D(r_1,r_2,P_1,P_3)},
  \end{equation}
  with
   \addtocounter{equation}{-1}
   \begin{subequations}
   \begin{align}
   C(r_1,r_2,P_1,P_3)=&  16\pi \left( {D - 3} \right)\left( {{P_1} - {P_3}} \right){{\left( {{r_1}{r_2}} \right)}^{D + 2}}\left( {{r_1^D}{r_2} - {r_1}{r_2^D}} \right),\\
      D(r_1,r_2,P_1,P_3)=& 2\left( {D - 1} \right){q^2}{r_1^3}{r_2^3}\left( { - {r_1^D}{r_2^3} + {r_1^3}{r_2^D}} \right)\nonumber\\
      & + \left( {D - 3} \right){\left( {{r_1}{r_2}} \right)^d}\left( { - 16\pi {P_1}{{\left( {{r_1}{r_2}} \right)}^2}\left( { - {r_1^D}{r_2} + {r_1}{r_2^D}} \right) - k\left( {D - 2} \right)\left( {D - 1} \right)\left( { - {r_1^D}{r_2^3} + {r_1^3}{r_2^D}} \right)} \right).
      \end{align}
   \end{subequations}
In what follows we study the efficiency of the ideal cycle in Fig. \ref{carnot} for the logarithmic $U(1)$ AdS black hole using the exact efficiency formula in high and low temperature domains. In order to verify the behaviour of efficiency as a function of $\beta$, we need to specify which parameters of the cycle remain constant as $\beta$ changes. There are many choices including the following two schemes as in \cite{Johnson2016a}. In the first scheme, the operating temperatures and pressures of the engine's ideal cycle ($T_1$, $T_2$, $P_1=P_2$ and $P_3=P_4$) in Fig. \ref{carnot} at points $1$ and $2$ are fixed. This is equivalent to specify the absorbed heat (Eq. \ref{heat}) along the isobar process $1\rightarrow2$ in Fig. \ref{carnot}}}. In the second scheme, the volumes and temperatures of the ideal cycle ($V_2$, $V_4$, $T_2\equiv T_H$ and $T_4\equiv T_C$) in Fig. \ref{carnot} at points $2$ and $4$ are specified and hold fixed. This corresponds to establish an engine with the initial and final volumes and temperatures in mind. In summary:

 \begin{itemize}
 	\item \textit{Scheme 1}:\\
 	 $(T_{1},T_{2})$ and $(P_{1},P_{3})$ $\to$ fixed ,\quad $T_{4}$ $\to$ found from the equation of state
 	\item \textit{Scheme 2}:\\
 	$(T_{2}, T_{4})$ and $(V_{2},V_{4})$ $\to$ fixed, \quad $(P_1, P_3)$ $\to$ found from the equation of state
 \end{itemize}

\begin{figure}[!htbp]
\begin{center}
	\epsfxsize=9 cm 
	\includegraphics[width=8 cm]{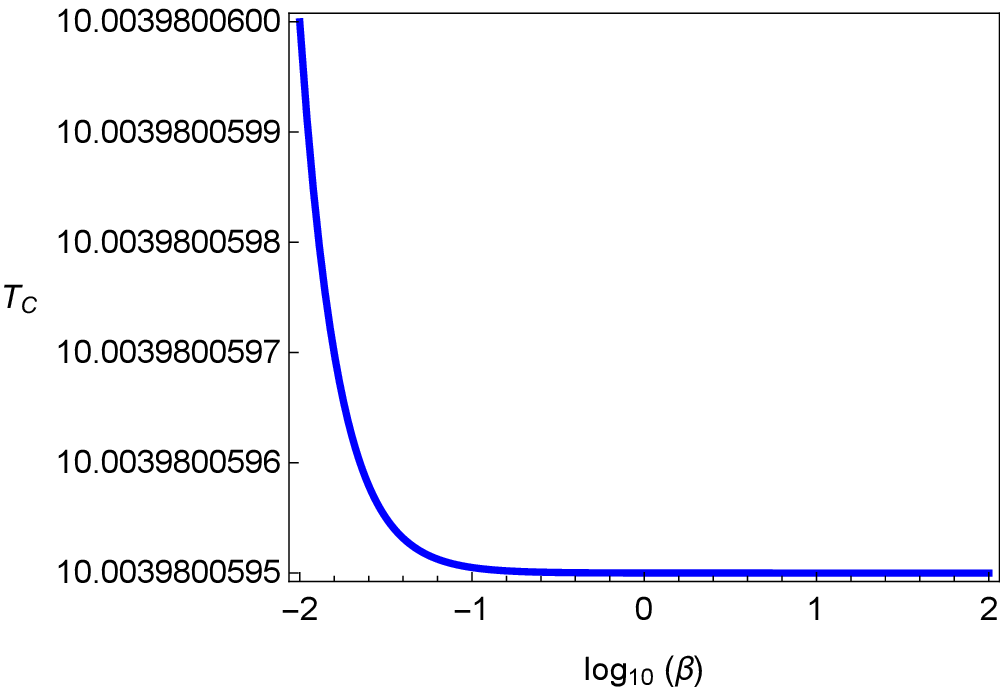}
	\hskip 1 cm
	\epsfxsize=9 cm 
    \includegraphics[width=8 cm]{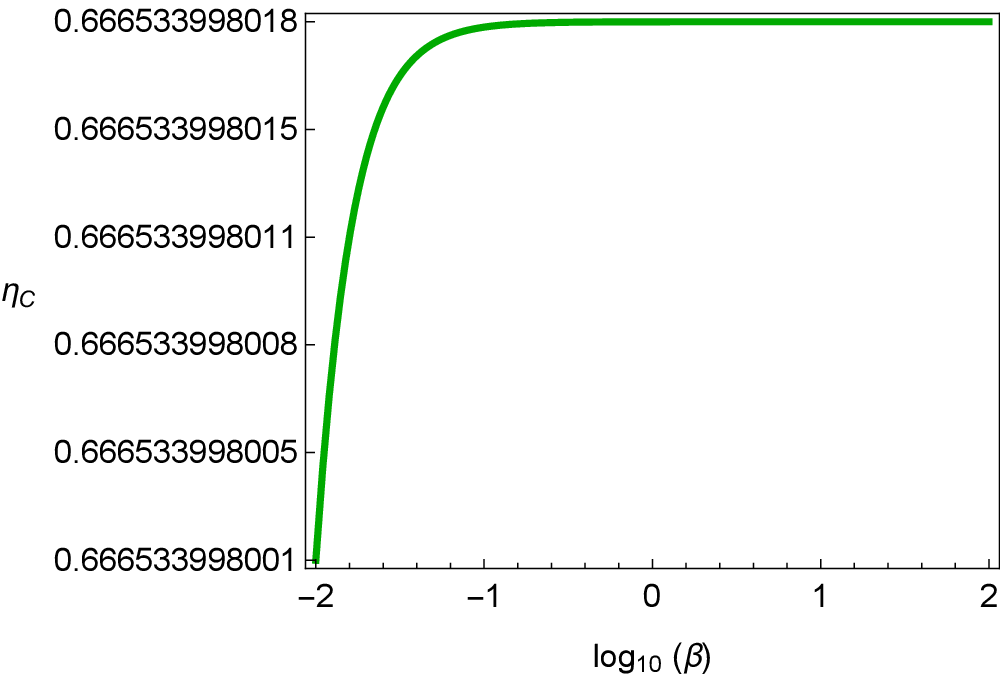}
	\caption{$T_C$ and $\eta_C$ versus log$_{10}(\beta)$ in scheme 1 for $D=4$, $T_1=20$, $T_2=30$, $P_1=1$, $P_4=0.5$ and $q=0.1$. }
	\label{scheme1_fig1}
	\end{center}
\end{figure}

\begin{figure}[!htbp]
\begin{center}
	  \epsfxsize=9 cm 
	\includegraphics[width=5.5 cm]{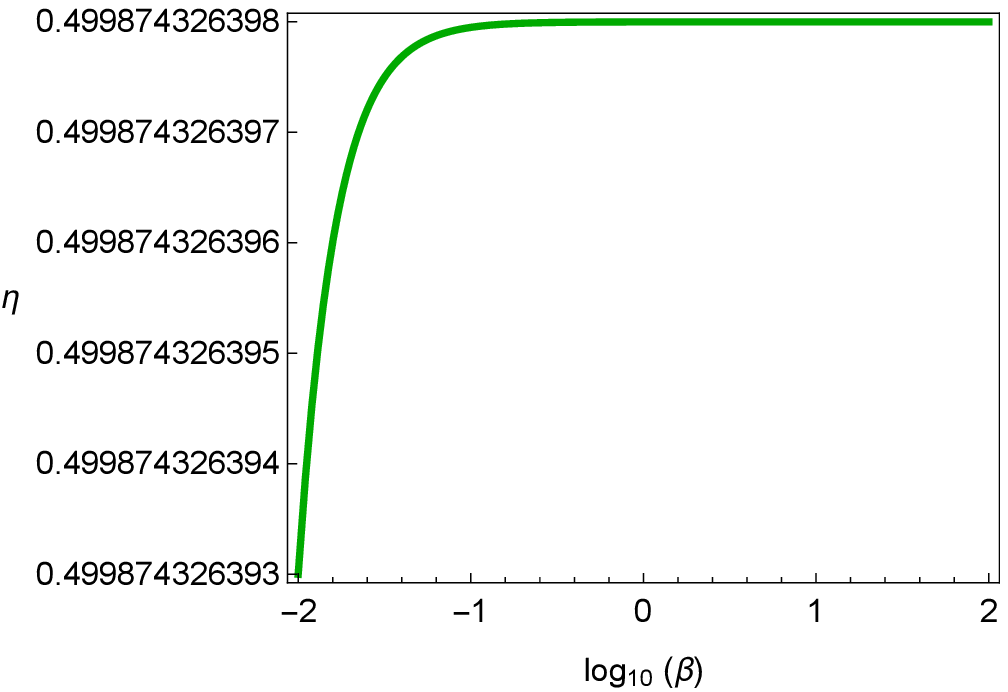}
	\hskip 0.5 cm
	\epsfxsize=9 cm 
	\includegraphics[width=5.5 cm]{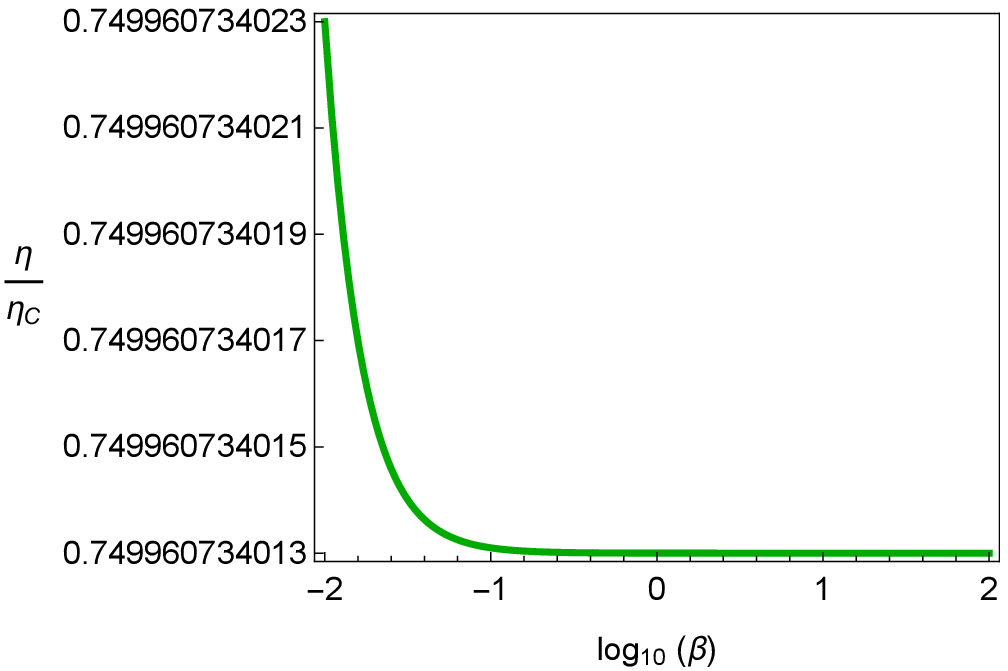}
	\hskip 0.5 cm
	\epsfxsize=9 cm 
    \includegraphics[width=5.5 cm]{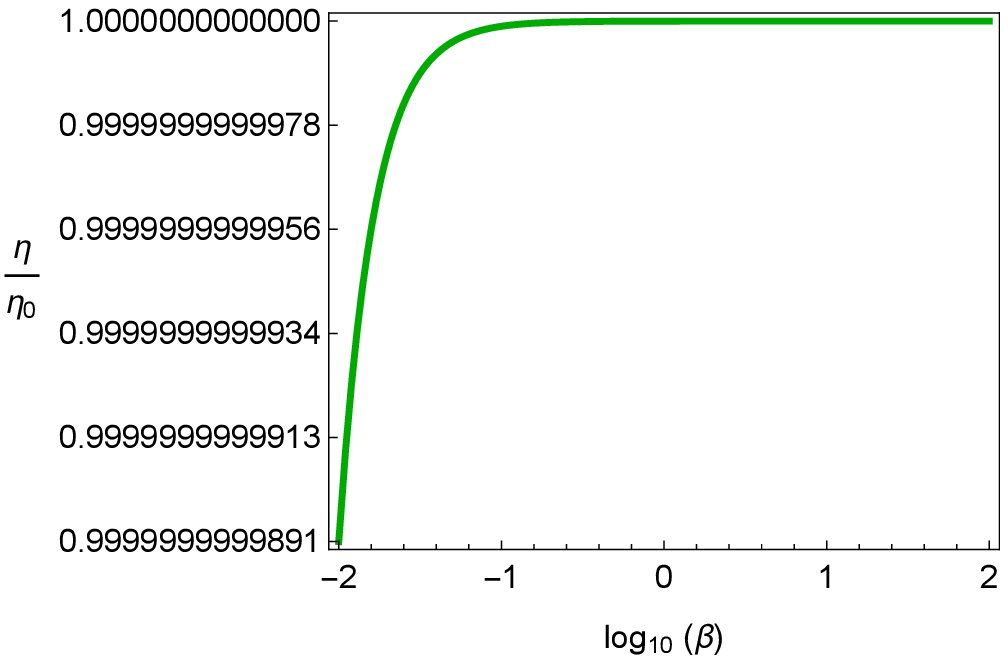}
	\caption{ $\eta$, $\eta_{C}$ and $\eta/\eta_{0}$ versus log$_{10}(\beta)$ in scheme 1 for $D=4$, $T_1=20$, $T_2=30$, $P_1=1$, $P_4=0.5$ and $q=0.1$.}
	\label{scheme1_fig2}
	\end{center}
\end{figure}

\textbf{High temperature domain:}
 As mentioned before, it is sensible to compare the efficiency of the desired engine cycle with the Carnot efficiency $\eta_C$, as well as the efficiency in the Einstein-Maxwell limit $\eta_0$, therefore we plot $\eta/\eta_C$ and $\eta/\eta_0$ as functions of $\beta$. It should be pointed out that, in scheme 1, $T_H\equiv T_2$ is specified but $T_C\equiv T_4$ is found from equation of state and hence changes with $\beta$. Fig. \ref{scheme1_fig1} shows $T_C$ and $\eta_C$ versus log$_{10}(\beta)$ for the range $0.01<\beta<100$ in this scheme. As $\beta$ goes to infinity that is a Maxwell limit (the limit for which our solutions approach the Reissner-Nordstr\"{o}m AdS black holes), $T_C$ decreases and $\eta_C$ increases. However the rate of changes with respect to $\beta$ is too slow for $\beta>0.1$. This feature is also seen for $\eta $, $\eta/\eta_C$ and $\eta/\eta_0$ in Fig. \ref{scheme1_fig2}.  Note that for the set of parameters in Figs. \ref{scheme1_fig1} and \ref{scheme1_fig2}, the black hole operates in the high temperature domain. These figures were also verified for $D=5$ and $D=6$ and the same behaviour was exhibited. One would expect the same behaviour for higher dimensional AdS spacetimes too.

\begin{figure}[!htbp]
\begin{center}
	\epsfxsize=9 cm 
	\includegraphics[width=5.5 cm]{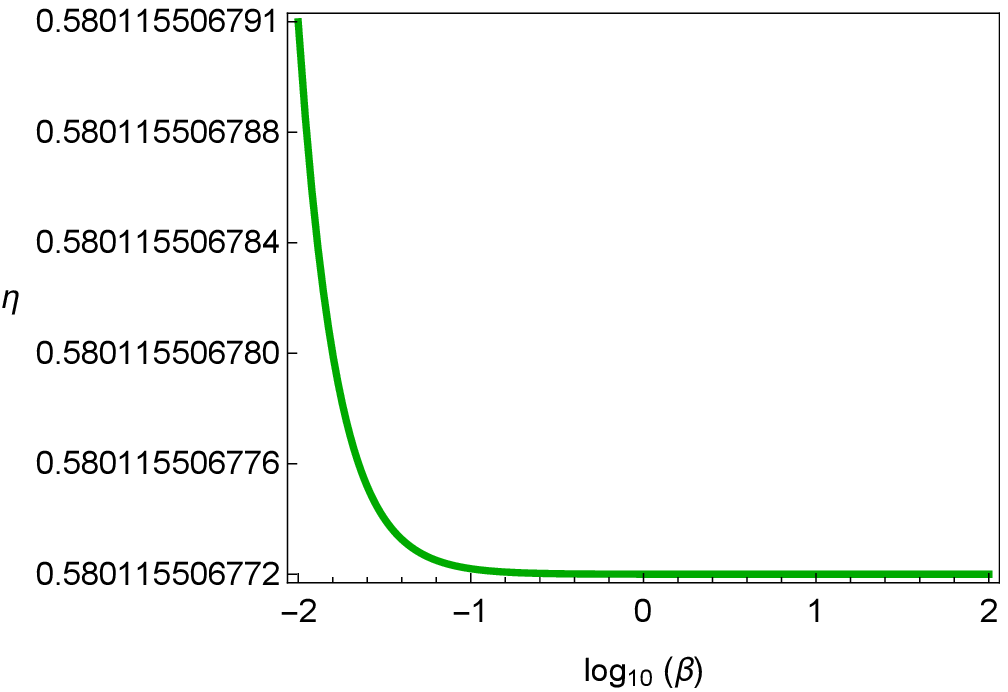}
	\hskip 0.5 cm
	\epsfxsize=9 cm 
	\includegraphics[width=5.5 cm]{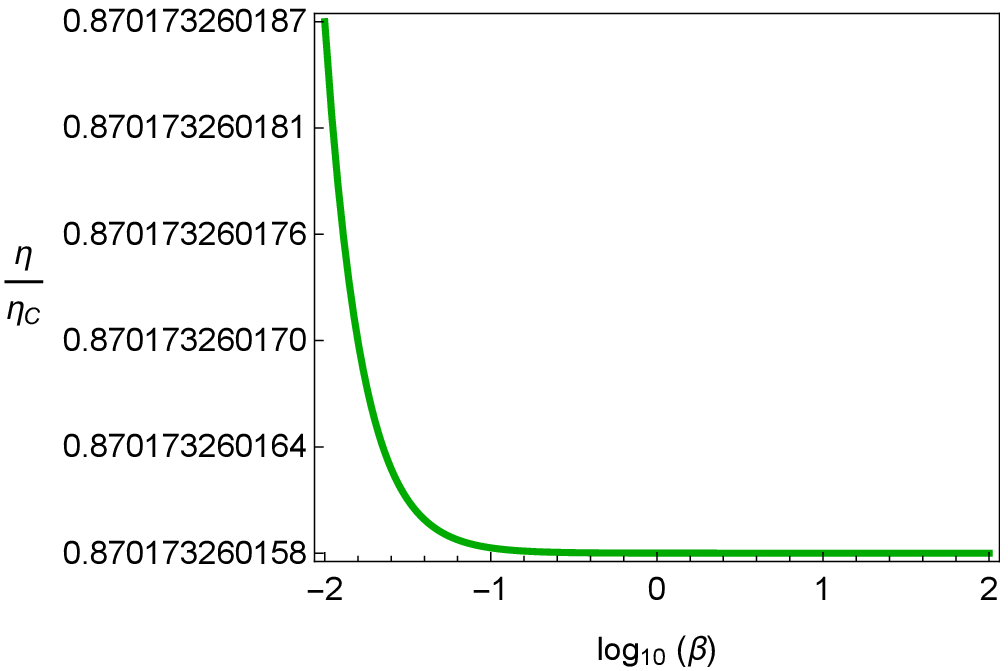}
	\hskip 0.5 cm
	\epsfxsize=9 cm 
    \includegraphics[width=5.5 cm]{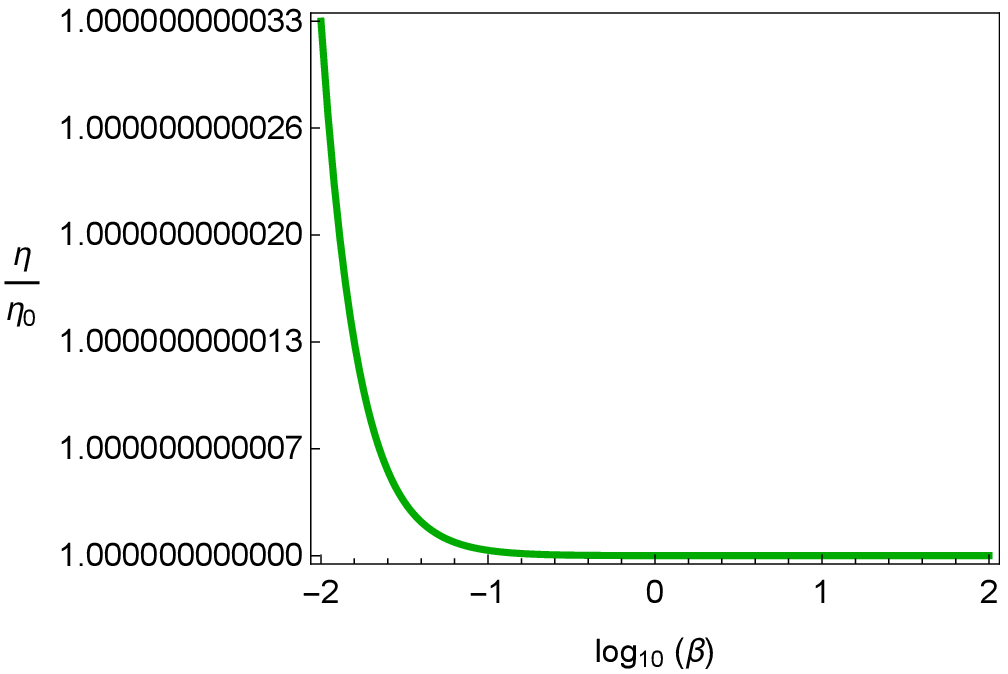}
	\caption{ $\eta$, $\eta/\eta_{C}$ and $\eta/\eta_{0}$ versus log$_{10}(\beta)$ in scheme 2 for $D=4$, $T_2=30$, $T_4=10$, $V_2=10000$, $V_4=5000$ and $q=0.1$. }
	\label{scheme2_fig}
	\end{center}
\end{figure}

While the Carnot efficiency $\eta_C$ is independent of $\beta$ in scheme 2, $P_1$, $P_3$ and $T_1$ which are evaluated from the equation of state vary with $\beta$. Fig. \ref{scheme2_fig} shows that contrary to scheme 1, in this scheme $\eta$ decreases with increasing $\beta$. The same qualitative behaviour as in scheme 1 is seen for $\eta/\eta_C$ and $\eta/\eta_0$ in this scheme. \\

\begin{figure}[!htbp]
\begin{center}
	 \epsfxsize=9 cm 
	\includegraphics[width=5.5 cm]{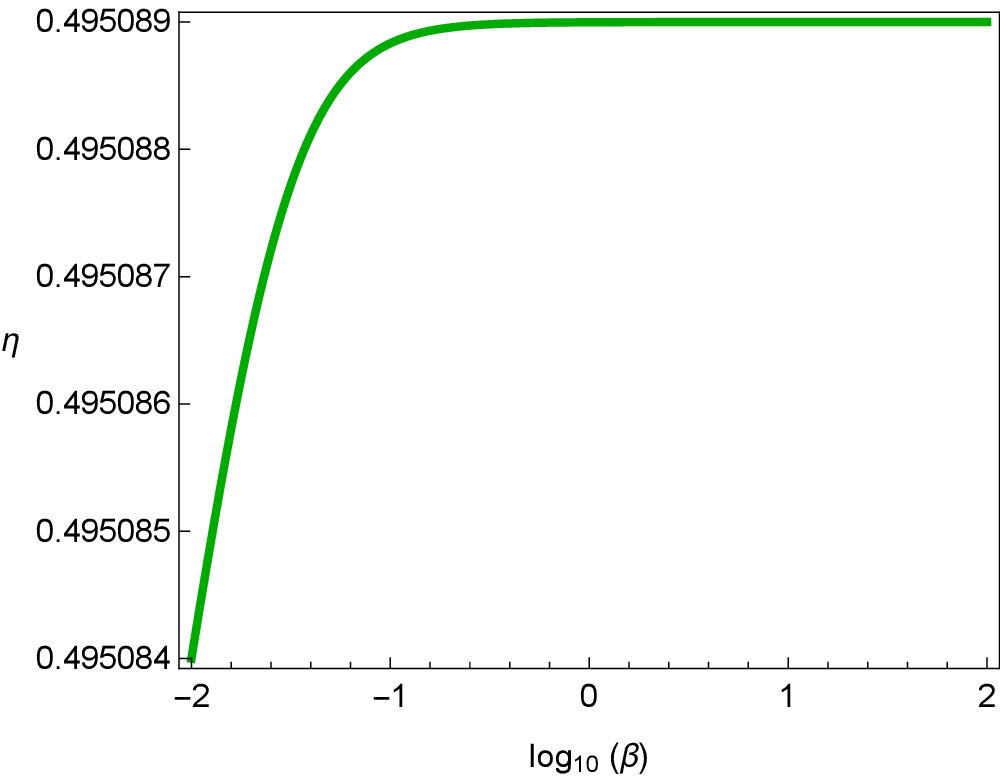}
	\hskip 0.5 cm
	\epsfxsize=9 cm 
	\includegraphics[width=5.5 cm]{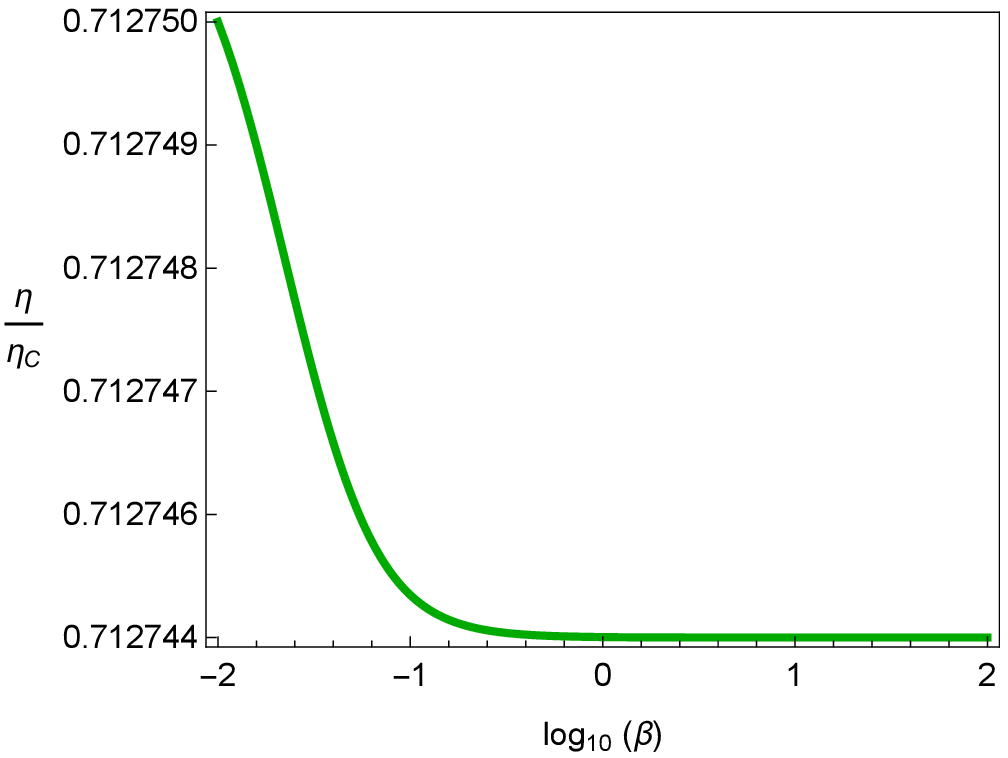}
	\hskip 0.5 cm
	\epsfxsize=9 cm 
	\includegraphics[width=5.5 cm]{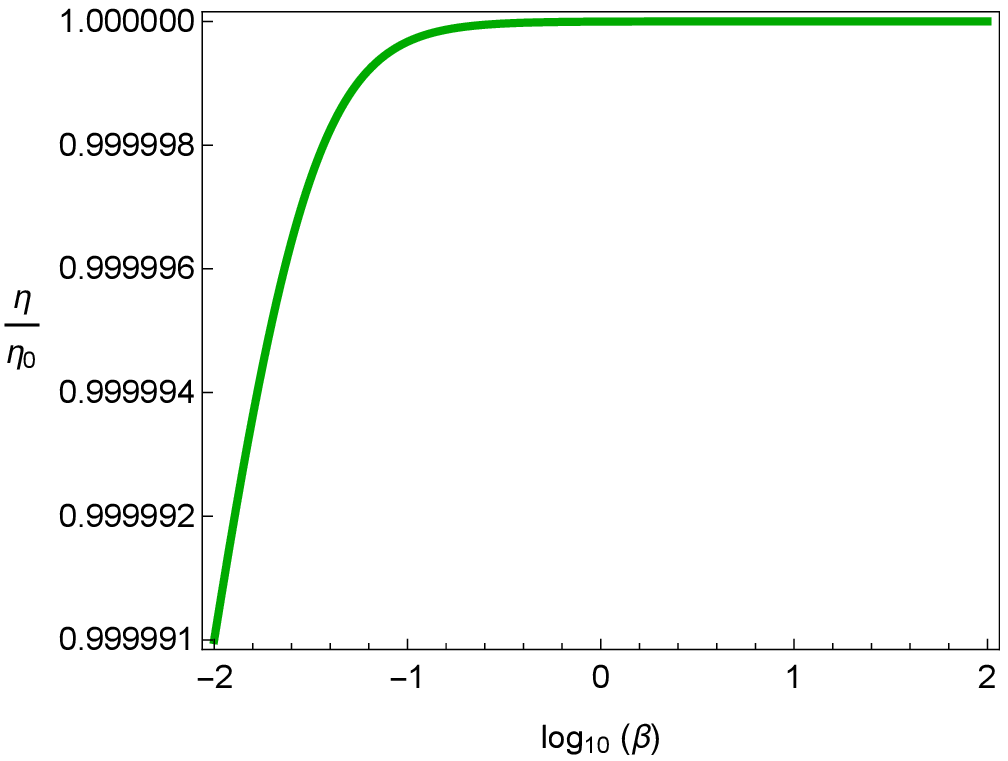}
	\caption{ $\eta$, $\eta_{C}$ and $\eta/\eta_{0}$ versus log$_{10}(\beta)$ in scheme 1 for $D=4$, $T_1=3$, $T_2=5$, $P_1=1$, $P_4=0.5$ and $q=0.1$. Here the critical temperature $T_{cr.}$ is around  $0.44$.  }
	\label{scheme1_LT}
	\end{center}
\end{figure}

\begin{figure}[!htbp]
\begin{center}
	 \epsfxsize=9 cm 
	\includegraphics[width=5.5 cm]{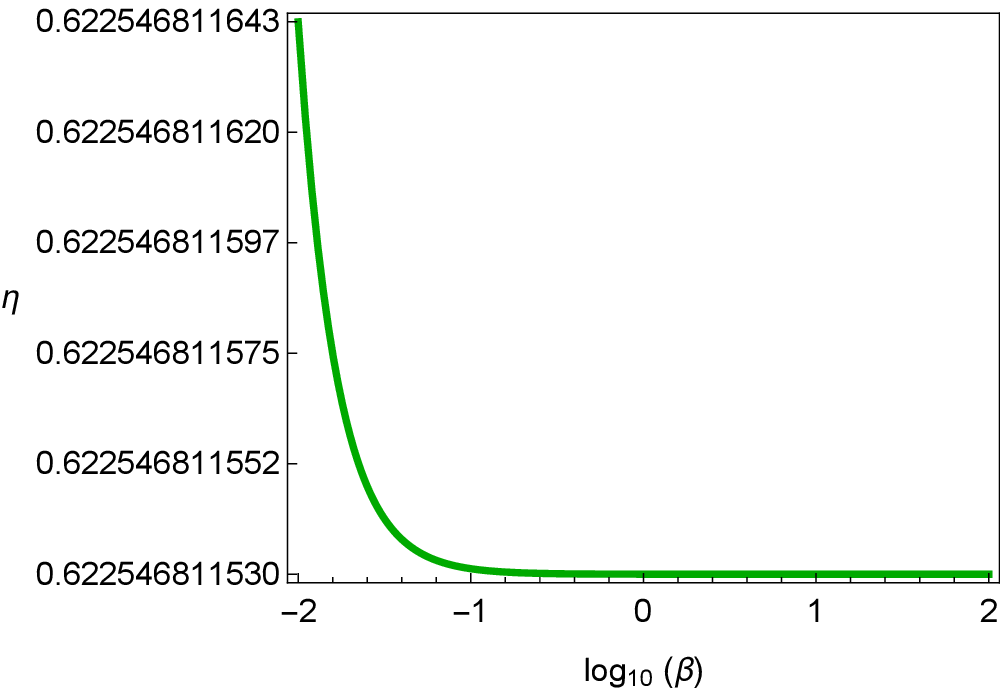}
	\hskip 0.5 cm
	\epsfxsize=9 cm 
	\includegraphics[width=5.5 cm]{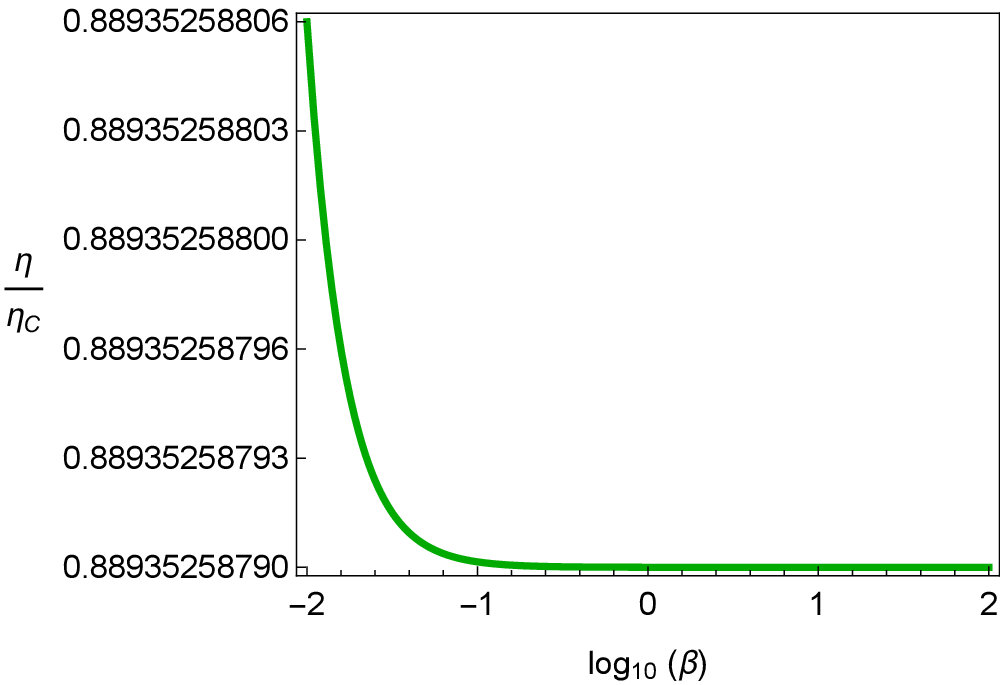}
	\hskip 0.5 cm
	\epsfxsize=9 cm 
	\includegraphics[width=5.5 cm]{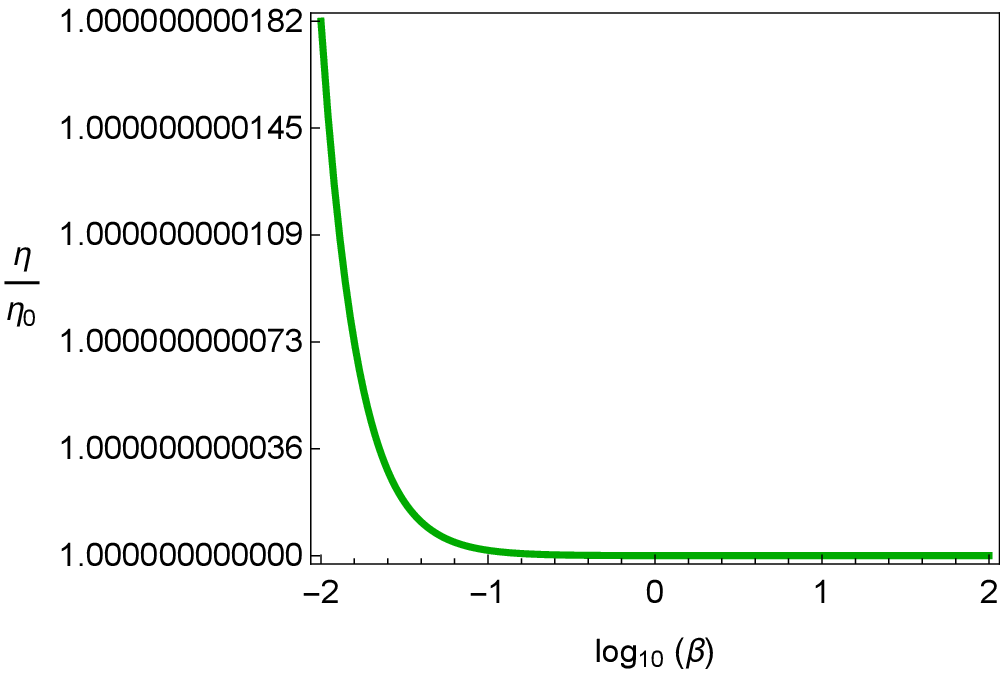}
	\caption{ $\eta$, $\eta/\eta_{C}$ and $\eta/\eta_{0}$ versus log$_{10}(\beta)$ in scheme 2 for $D=4$, $T_2=5$, $T_4=1.5$, $V_2=10000$, $V_4=5000$ and $q=0.1$. Here the critical temperature $T_{cr.}$ is around  $0.44$ . }
	\label{scheme2_LT}
	\end{center}
\end{figure}

\textbf{Low temperature domain:}
If we let the black hole engine to work in a low temperature domain (but still higher than critical temperature), i.e., decreasing both $T_C$ and $T_H$, and examine the efficiency of the ideal cycle for the schemes introduced before, it is seen that $\eta$, $\eta/\eta_C$ and $\eta/\eta_0$ behave the same way as the high temperature domain (See Figs. \ref{scheme1_LT} and \ref{scheme2_LT}). Comparing $\eta$ in Figs. \ref{scheme1_fig2} and \ref{scheme1_LT}, it is clear that for the set of parameter data we have chosen (keeping pressure constant while decreasing temperature) the efficiency decreases in scheme 1. However, for scheme 2, the efficiency increases by decreasing temperature while keeping the volumes constant (compare Figs. \ref{scheme2_fig}) and \ref{scheme2_LT}.) 

\subsection{The engine's efficiency in weak and strong coupling regimes}\label{secht}

There are four important cases of physical interest that can be analytically  approximated: High and low temperature regions with weak and strong couplings. In order to analyze these regions we need to distinguish between diverse approximations. Although the expansions of electric field (\ref{electric field}), ADM mass (\ref{MADM}), temperature (\ref{temp}) and vacuum polarization (\ref{vacuum polarization}) around the large values of $r_+$ and around $\beta \to \infty$ are the same, but their meanings are different. Expansion around the large values of $r_+$ means we are far from the charged source (here, the black hole's singularity), and we can freely vary the non-linearity parameter $\beta$ for both the weak and strong coupling regimes. But, expansion around the weak coupling limit ($\beta \to \infty$) restricts us to vary $\beta$ in the weak coupling regime, no matter how far we are from the source. This distinction is vital for studying efficiency of holographic heat engines in this section.

 For now, we will analyze these domains for engines with spherical black holes as working substance. Results will be generalized to the holographic heat engines that are planar or hyperbolic in topology in Sect. \ref{planar-hyperbolic engines}.\\
 \par  
 \textbf{ High temperature with weak couplings ($T_{\rm{high}}, \beta_{\rm{large}}$):} Since the enthalpy ($H$) is a function of $S$, $P$ and $Q$ (it is not an explicit function of $T$), we cannot simply use the exact efficiency formula to obtain the high temperature limit. However, due to asymptotic behaviour of AdS black hole, it can analytically be approximated. The reason is that, according to Eqs. (\ref{temp}) and (\ref{eos_HT}), the AdS black holes with a fixed pressure in the high temperature limit behaves as $T \propto {r_+}$ (corresponding to the large black hole region) and one concludes that the event horizon radius should be large. Using this fact, it is inferred that the expansion of $\eta$ around large values of $r_+$ leads to the analytic efficiency formula in the high temperature limit. This expansion works for both weak and strong couplings.
The series expansion for efficiency about $r_+=\infty$ can be achieved by replacing $M|_{\beta=\infty}$ (\ref{massinf}) in the exact efficiency formula \ref{EEF} or making use of Eqs. (\ref{work}) and (\ref{heat}) both expanded in the high temperature limit or equivalently about $r_+=\infty$. The latter produces more accurate estimates when compared to the results obtained from the exact efficiency formula. Accordingly, we expand Eq. (\ref{temp}) about $r_+=\infty$ and solve it to find $r_+$  as a function of $T$, which for $D=4$ yields
\begin{eqnarray}\label{rp4}
r_+&=&\frac{T}{{2P}} - \frac{1}{{4\pi T}} + \frac{{P\left( {8\pi P{q^2} - 1} \right)}}{{8{\pi ^2}{T^3}}} + \frac{{{P^2}\left( {16\pi P{q^2} - 1} \right)}}{{8{\pi ^3}{T^5}}} - \frac{{{P^3}\left( {5{\beta ^2} - 120\pi P{q^2}{\beta ^2} + 192{\pi ^2}{P^2}{q^4}{\beta ^2} + 64{\pi ^3}{P^3}{q^4}} \right)}}{{32{\pi ^4}{\beta ^2}{T^7}}}\nonumber\\
&& - \frac{{{P^4}\left( {7{\beta ^2} - 224\pi P{q^2}{\beta ^2} + 896{\pi ^2}{P^2}{q^4}{\beta ^2} + 256{\pi ^3}{P^3}{q^4}} \right)}}{{32{\pi ^5}{\beta ^2}{T^9}}} + O{\left[ {\frac{1}{T}} \right]^{11}}.
\end{eqnarray}
In $D=5$, it is given by
\begin{eqnarray}\label{rp5}
r_+&=&\frac{{3T}}{{4P}} - \frac{1}{{2\pi T}} - \frac{P}{{3{\pi ^2}{T^3}}} + \frac{{4{P^2}\left( {32{\pi ^2}{P^2}{q^2} - 27} \right)}}{{243{\pi ^3}{T^5}}} + \frac{{4{P^3}\left( {128{\pi ^2}{P^2}{q^2} - 45} \right)}}{{243{\pi ^4}{T^7}}} + \frac{{112{P^4}\left( {128{\pi ^2}{P^2}{q^2} - 27} \right)}}{{2187{\pi ^5}{T^9}}}\nonumber\\
&& - \frac{{32{P^5}\left( {15309{\beta ^2} - 103680{\pi ^2}{P^2}{q^2}{\beta ^2} + 10240{\pi ^4}{P^4}{q^4}{\beta ^2} + 2048{\pi ^5}{P^5}{q^4}} \right)}}{{177147{\pi ^6}{\beta ^2}{T^{11}}}} + O{\left[ {\frac{1}{T}} \right]^{13}},
\end{eqnarray}
and in $D=6$
\begin{eqnarray}\label{rp6}
r_+&=&
\frac{T}{P} - \frac{3}{{4\pi T}} - \frac{{9P}}{{16{\pi ^2}{T^3}}} - \frac{{27{P^2}}}{{32{\pi ^3}{T^5}}} + \frac{{{P^3}\left( {32{\pi ^3}{P^3}{q^2} - 405} \right)}}{{256{\pi ^4}{T^7}}} + \frac{{3{P^4}\left( {128{\pi ^3}{P^3}{q^2} - 567} \right)}}{{512{\pi ^5}{T^9}}}+ \frac{{81{P^5}\left( {80{\pi ^3}{P^3}{q^2} - 189} \right)}}{{2048{\pi ^6}{T^{11}}}}\nonumber\\
&& + \frac{{297{P^6}\left( {160{\pi ^3}{P^3}{q^2} - 243} \right)}}{{4096{\pi ^7}{T^{13}}}} - \frac{{{P^7}\left( {2814669{\beta ^2} - 2594592{\pi ^3}{P^3}{q^2}{\beta ^2} + 7168{\pi ^6}{P^6}{q^4}{\beta ^2} + 1024{\pi ^7}{P^7}{q^4}} \right)}}{{65536{\pi ^8}{\beta ^2}{T^{15}}}} + O{\left[ {\frac{1}{T}} \right]^{17}}.\nonumber\\
\end{eqnarray}
Using Eqs. (\ref{rp4})-(\ref{rp6}), the thermodynamic volume of the $D$ dimensional spherical black hole can easily be found using the relation $V=\frac{\Sigma_{D-2}}{D-1}r^{D-1}$ and, when inserted in Eq. (\ref{work}), gives the work in the high temperature limit. Substituting Eqs. (\ref{rp4})-(\ref{rp6}) in Eq. (\ref{specific}) and integrating with respect to $T$, the absorbing heat in $D=4$ is given by 
\begin{eqnarray}
{Q_H} &=& \frac{{\pi {T^3}}}{{6{P^2}}} + \frac{{16\pi P{q^2} - 1}}{{8\pi T}} + \frac{{P\left( {24\pi P{q^2} - 1} \right)}}{{12{\pi ^2}{T^3}}} - \frac{{3{P^2}\left( {5{\beta ^2} - 160\pi P{q^2}{\beta ^2} + 320{\pi ^2}{P^2}{q^4}{\beta ^2} + 128{\pi ^3}{P^3}{q^4}} \right)}}{{160{\pi ^3}{\beta ^2}{T^5}}}\nonumber\\
 &&- \frac{{{P^3}\left( {{\beta ^2} - 40\pi P{q^2}{\beta ^2} + 192{\pi ^2}{P^2}{q^4}{\beta ^2} + 64{\pi ^3}{P^3}{q^4}} \right)}}{{8{\pi ^4}{\beta ^2}{T^7}}} + \left. {O{{\left[ {\frac{1}{T}} \right]}^9}} \right|_{_{{T_1}}}^{{T_2}}.
\end{eqnarray}
In $D=5$, we have
\begin{eqnarray}
{Q_H} &=& \frac{{81{\pi ^2}{T^4}}}{{512{P^3}}} - \frac{{27\pi {T^2}}}{{128{P^2}}} + \frac{{64{\pi ^2}{P^2}{q^2} - 9}}{{96\pi {T^2}}} + \frac{{5P\left( {256{\pi ^2}{P^2}{q^2} - 27} \right)}}{{864{\pi ^2}{T^4}}} + \frac{{7{P^2}\left( {320{\pi ^2}{P^2}{q^2} - 27} \right)}}{{648{\pi ^3}{T^6}}}\nonumber\\
&& - \frac{{{P^3}\left( {1701{\beta ^2} - 24192{\pi ^2}{P^2}{q^2}{\beta ^2} + 4096{\pi ^4}{P^4}{q^4}{\beta ^2} + 1024{\pi ^5}{P^5}{q^4}} \right)}}{{2916{\pi ^4}{\beta ^2}{T^8}}} + \left. {O{{\left[ {\frac{1}{T}} \right]}^{10}}} \right|_{_{{T_1}}}^{{T_2}},
\end{eqnarray}
and in $D=6$
\begin{eqnarray}
{Q_H} &=& \frac{{8{\pi ^2}{T^5}}}{{15{P^4}}} - \frac{{4\pi {T^3}}}{{3{P^3}}} + \frac{{128{P^3}{\pi ^3}{q^2} - 81}}{{288{\pi ^2}{T^3}}} + \frac{{3P\left( {160{\pi ^3}{P^3}{q^2} - 81} \right)}}{{320{\pi ^3}{T^5}}} + \frac{{9{P^2}\left( {64{\pi ^3}{P^3}{q^2} - 27} \right)}}{{128{\pi ^4}{T^7}}}\nonumber\\
&& + \frac{{15{P^3}\left( {224{\pi ^3}{P^3}{q^2} - 81} \right)}}{{256{\pi ^5}{T^9}}} - \frac{{{P^4}\left( {1082565{\beta ^2} - 3421440{\pi ^3}{P^3}{q^2}{\beta ^2} + 22528{\pi ^6}{P^6}{q^4}{\beta ^2} + 4096{\pi ^7}{P^7}{q^4}} \right)}}{{90112{\pi ^6}{T^{11}}{\beta ^2}}}\nonumber\\
 &&+ \left. {O{{\left[ {\frac{1}{T}} \right]}^{13}}} \right|_{_{{T_1}}}^{{T_2}}.
\end{eqnarray}
With the work and absorbed heat relations in hand, finally the efficiency of the logarithmic $U(1)$ AdS black hole engine in the high temperature limit in $D$ dimensional spacetime is achieved by using $\eta=W/Q_H$. Looking at above expansions, it is seen that the non-linearity parameter $\beta$ appears in the work and heat relations at order $T^{-5}$ for $D=4$, at order  $T^{-8}$ for $D=5$ and at order $T^{-11}$ for $D=6$. Hence, the contributions of these terms which contain $\beta$ factor are completely negligible in the weak couplings. However, for the strong couplings ($\beta \rightarrow 0$), the effect of these terms are sensible.

\begin{figure}[!htbp]
\begin{center}
	\epsfxsize=9 cm 
	\includegraphics[width=5.5 cm]{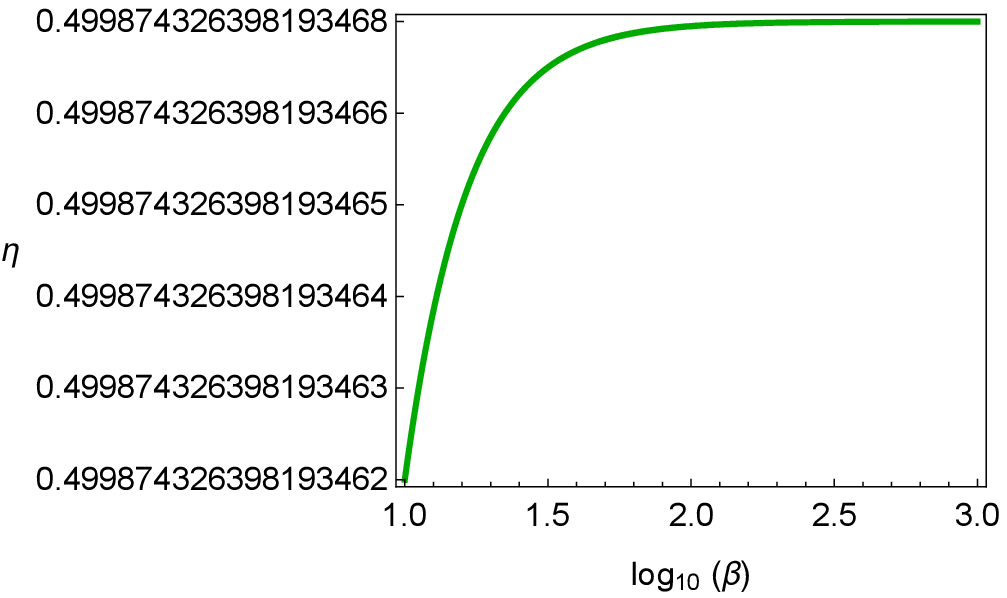}
	\hskip 0.5 cm
	\epsfxsize=9 cm 
	\includegraphics[width=5.5 cm]{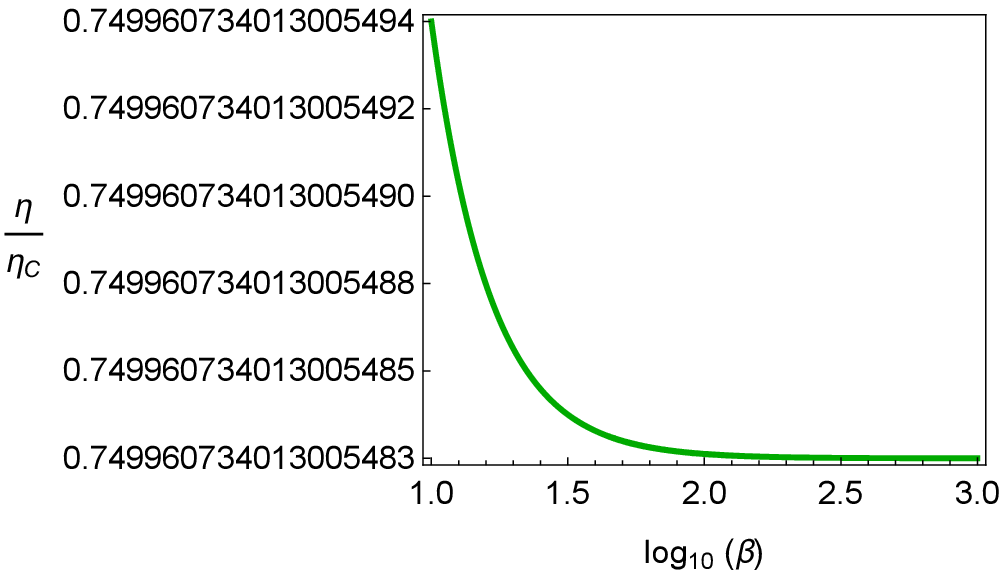}
	\hskip 0.5 cm
	\epsfxsize=9 cm 
    \includegraphics[width=5.5 cm]{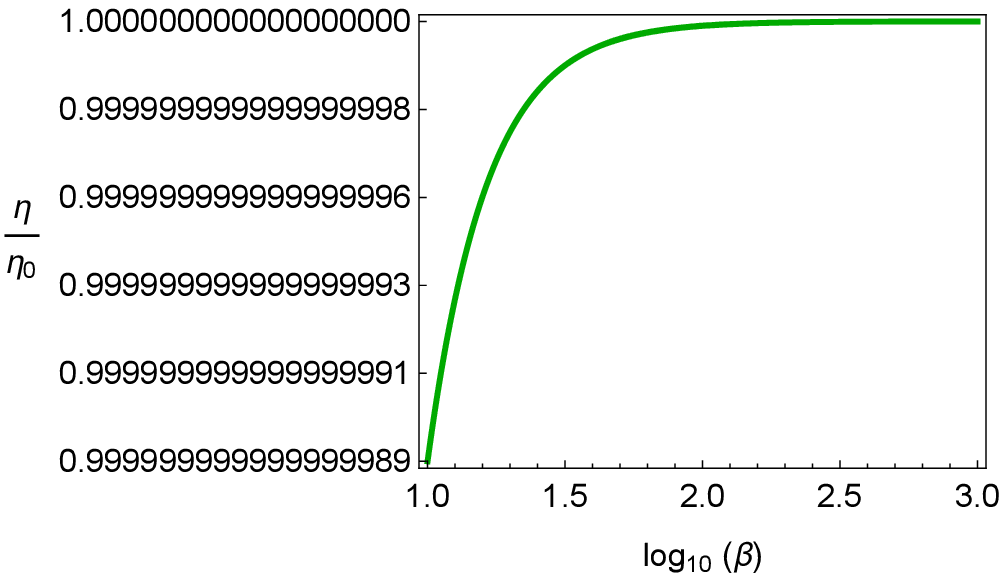}
	\caption{ $\eta$, $\eta/\eta_{C}$ and $\eta/\eta_{0}$ versus log$_{10}(\beta)$ in scheme 1 in high temperature domain with weak couplings for $D=4$, $T_1=20$, $T_2=30$, $P_1=1$, $P_4=0.5$ and $q=0.1$ .}
	\label{HT_WC}
	\end{center}
\end{figure}

 If $\eta$, $\eta/\eta_{C}$ and $\eta/\eta_{0}$ are plotted as functions of ${\rm{log}}_{10}(\beta)$ in the high temperature domain with weak couplings (in the range $10<\beta<1000$), the corresponding diagrams like those of Fig. \ref{scheme1_fig2} are obtained (See Fig. \ref{HT_WC}): $\eta$  and $\eta/\eta_{0}$ follow the strictly increasing functional pattern (like the function$f(x)=1/x$) and $\eta/\eta_{C}$ is a strictly decreasing function (like the function $f(x)=\sqrt{x}$). By that we mean when $\beta$ runs from strong to weak coupling region, it has a monotonically smooth behaviour. Plotting $\eta$, $\eta/\eta_{C}$ and $\eta/\eta_{0}$ in scheme 2 in high temperature region with weak couplings also reveal the similar behaviour to those of Fig. (\ref{scheme2_fig}). 

\begin{figure}[!htbp]
\begin{center}
	\epsfxsize=9 cm 
	\includegraphics[width=5.5 cm]{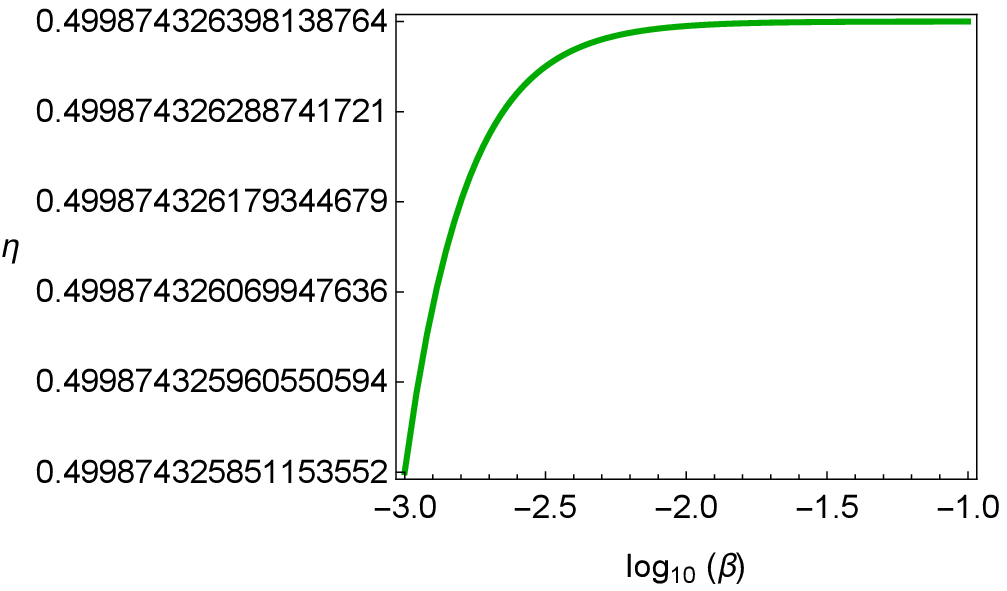}
	\hskip 0.5 cm
	\epsfxsize=9 cm 
	\includegraphics[width=5.5 cm]{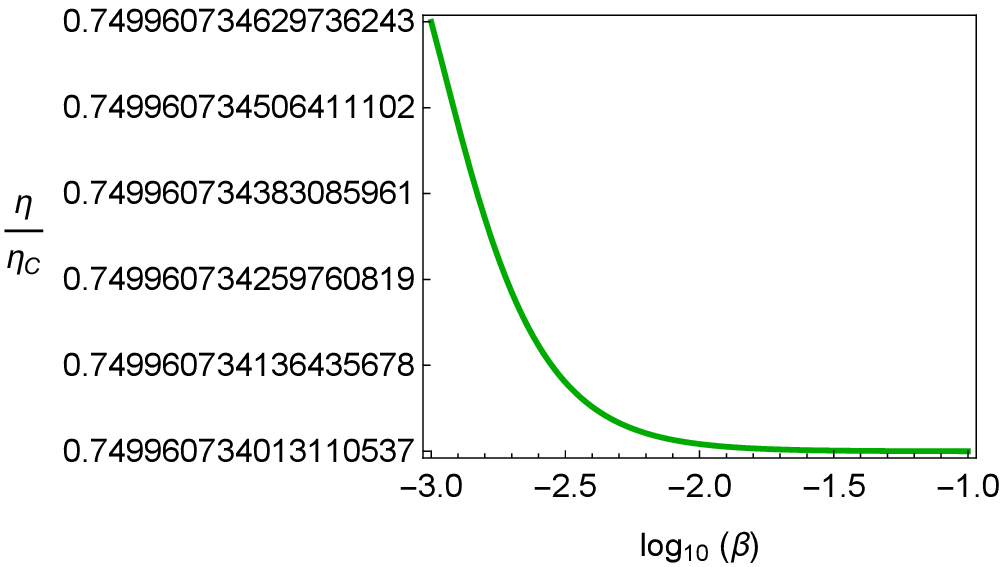}
	\hskip 0.5 cm
	\epsfxsize=9 cm 
    \includegraphics[width=5.5 cm]{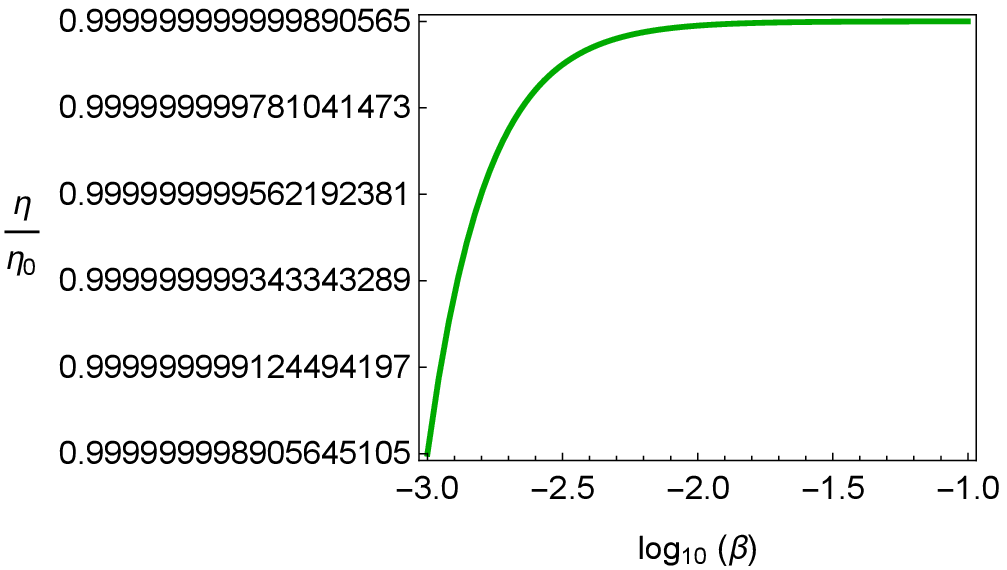}
	\caption{ $\eta$, $\eta/\eta_{C}$ and $\eta/\eta_{0}$ versus log$_{10}(\beta)$ in scheme 1 in high temperature domain with strong couplings for $D=4$, $T_1=20$, $T_2=30$, $P_1=1$, $P_4=0.5$ and $q=0.1$. }
	\label{HT_SC}
	\end{center}
\end{figure}

\textbf{ High temperature with strong couplings ($T_{\rm{high}}, \beta_{\rm{small}} $):} Here, we apply the method we used for evaluating efficiency in the high temperature limit with weak couplings ($\beta \rightarrow \infty$) again, but this time to strong couplings ($\beta \rightarrow 0$). The results for $\eta$, $\eta/\eta_{C}$ and $\eta/\eta_{0}$ are plotted for a range of  $10^{-3}<\beta<10^{-1}$ in Fig. \ref{HT_SC} for scheme 1. The qualitative similarity between these diagrams and those in Figs. \ref{scheme1_fig2} and \ref{HT_WC} emphasizes the smooth behaviour of efficiency in the range of $0<\beta<\infty$.

\begin{figure}[!htbp]
\begin{center}
	\epsfxsize=9 cm 
	\includegraphics[width=5.5 cm]{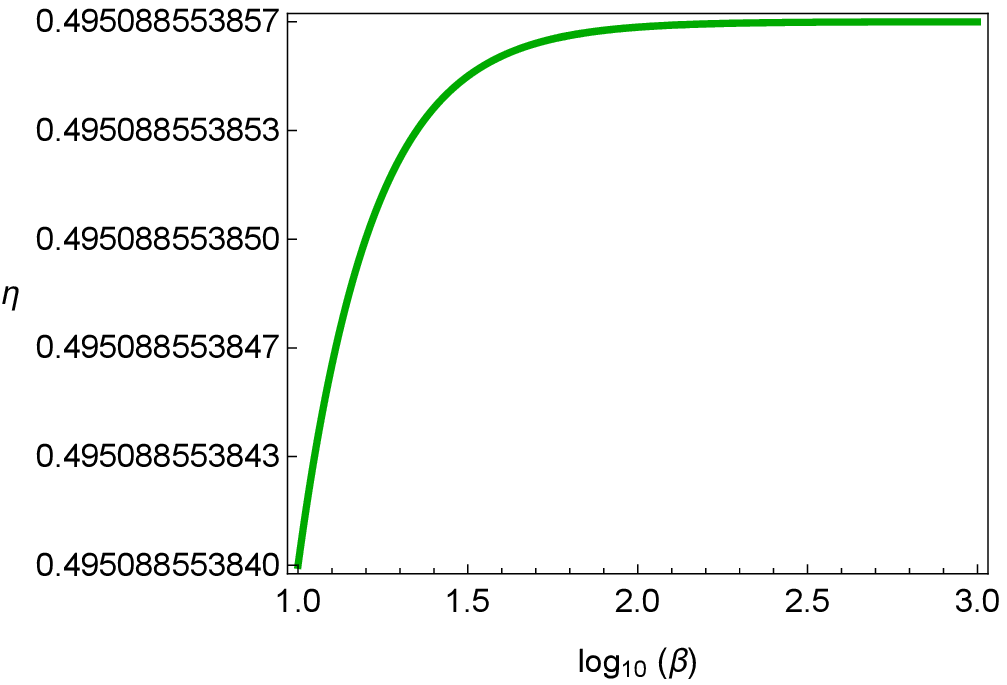}
	\hskip 0.5 cm
	\epsfxsize=9 cm 
	\includegraphics[width=5.5 cm]{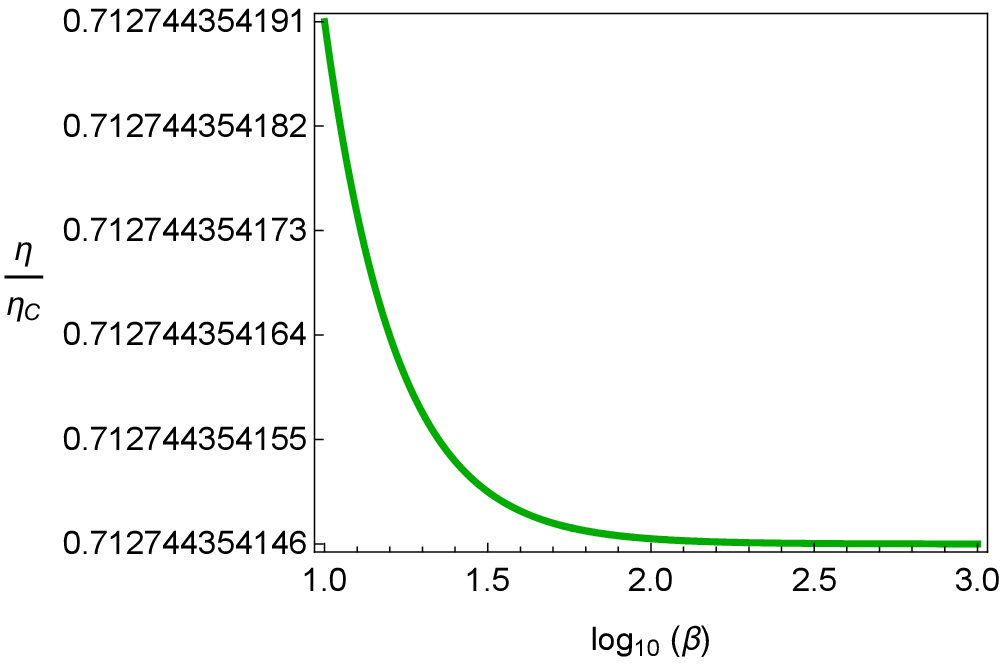}
	\hskip 0.5 cm
	\epsfxsize=9 cm 
    \includegraphics[width=5.5 cm]{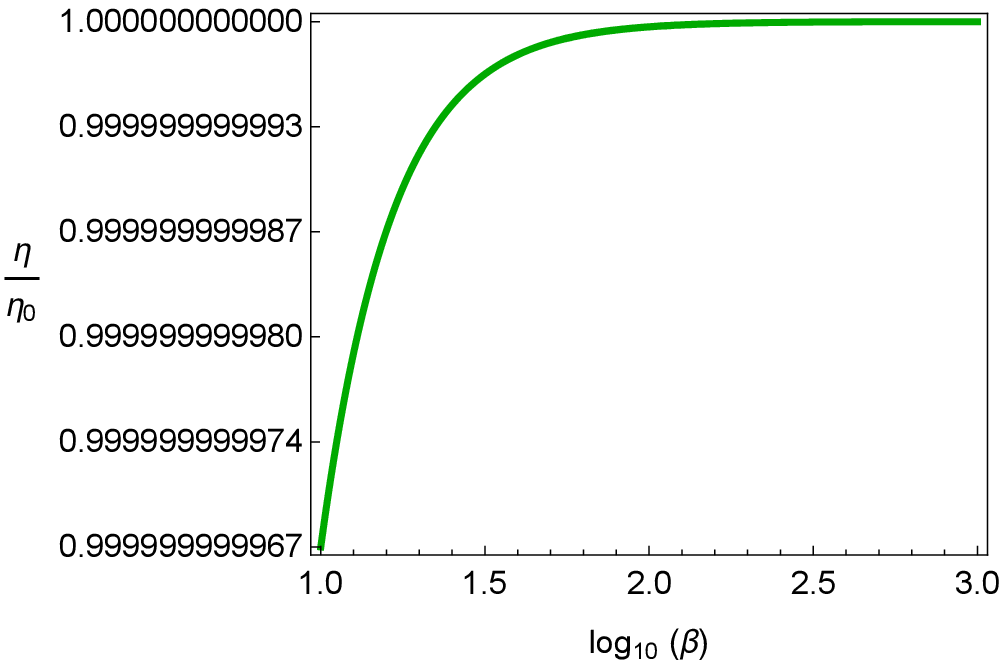}
	\caption{ $\eta$, $\eta/\eta_{C}$ and $\eta/\eta_{0}$ versus log$_{10}(\beta)$ in scheme 1 in low temperature domain with weak couplings for $D=4$, $T_1=3$, $T_2=5$, $P_1=1$, $P_4=0.5$ and $q=0.1$. }
	\label{LT_WC}
	\end{center}
\end{figure}

\textbf{ Low temperature with weak couplings ($T_{\rm{low}}, \beta_{\rm{large}} $)}: In the low temperature limit $T$ is no longer proportional to $r_{+}$ and so expanding the efficiency about $T=0$ is not possible. Instead, we run its series expansion about $\beta=\infty$ in low temperature limit. The relation for $\eta|_{\beta=\infty}$  was already derived in Eq. (\ref{effbetainf}). Using this equation $\eta$, $\eta/\eta_{C}$ and $\eta/\eta_{0}$ are sketched for low temperature domain (higher than critical temperature) in Fig. \ref{LT_WC} for scheme 1. As expected, a qualitative agreement is seen with Fig. \ref{scheme1_LT}.

\begin{figure}[!htbp]
\begin{center}
	\epsfxsize=9 cm 
	\includegraphics[width=5.5 cm]{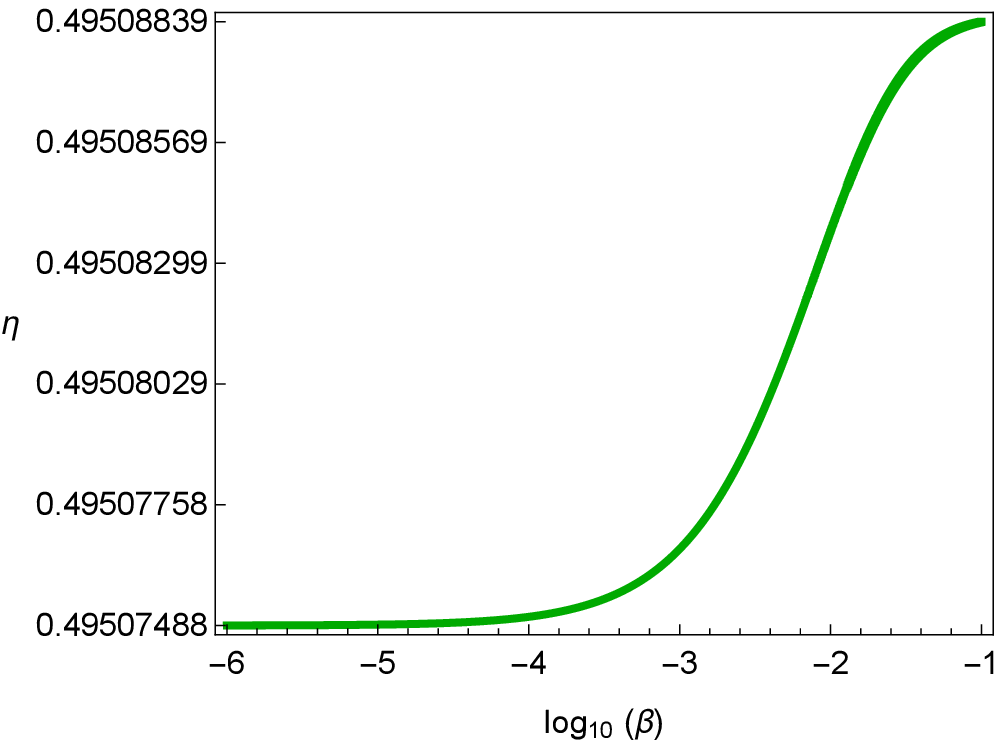}
	\hskip 0.5 cm
	\epsfxsize=9 cm 
	\includegraphics[width=5.5 cm]{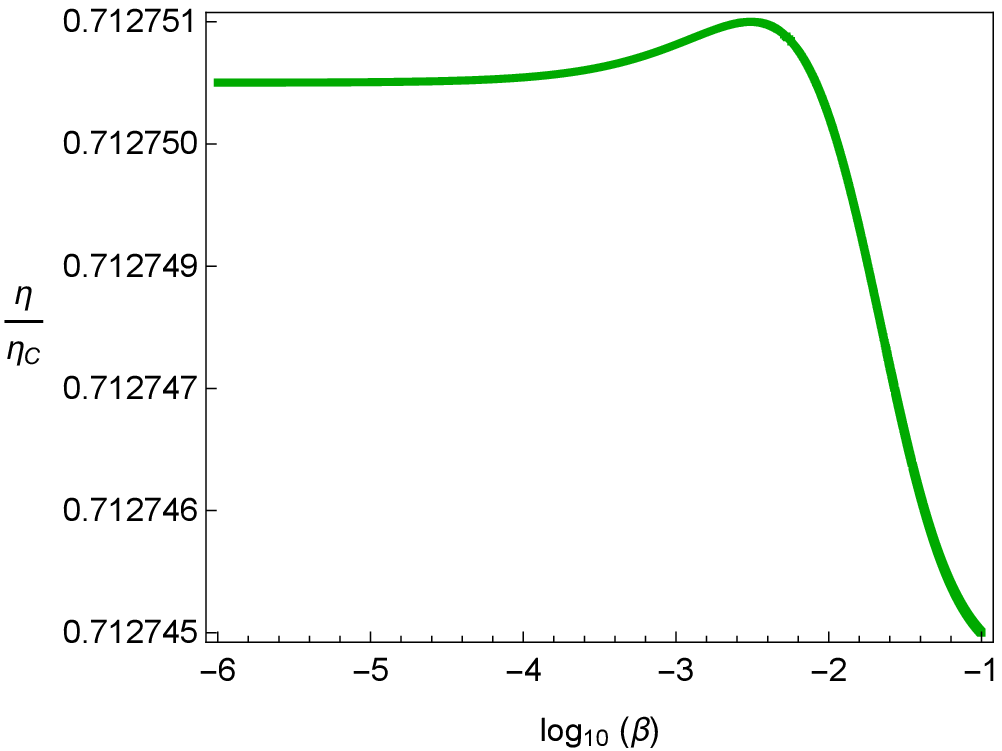}
	\hskip 0.5 cm
	\epsfxsize=9 cm 
    \includegraphics[width=5.5 cm]{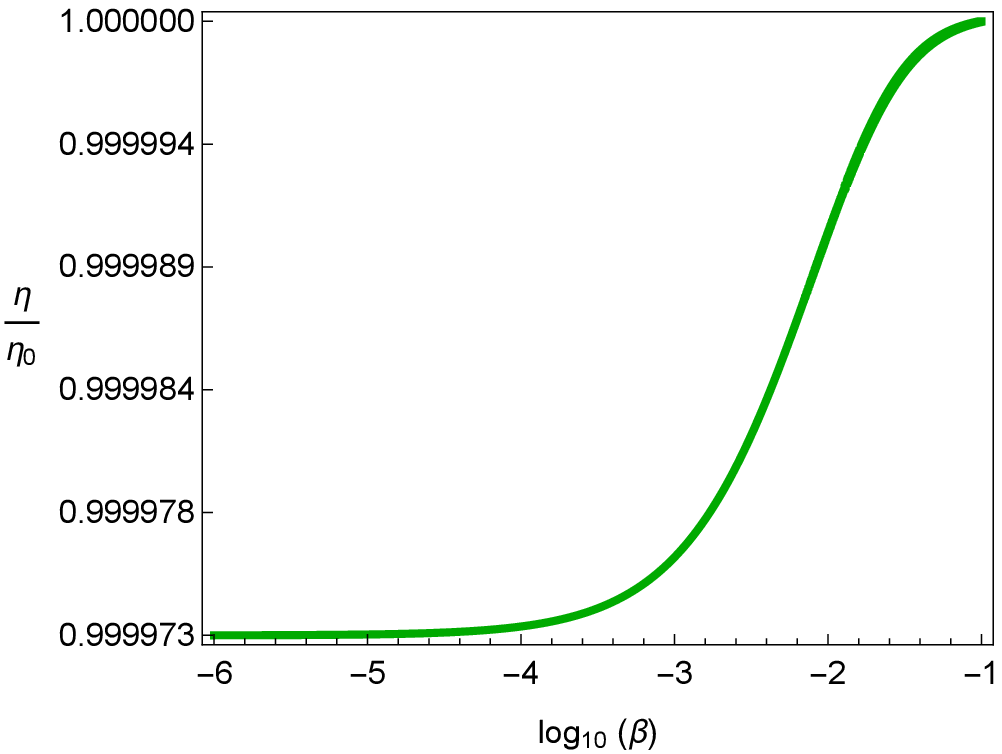}
	\caption{ $\eta$, $\eta/\eta_{C}$ and $\eta/\eta_{0}$ versus log$_{10}(\beta)$ in scheme 1 in low temperature domain with strong couplings for $D=4$, $T_1=3$, $T_2=5$, $P_1=1$, $P_4=0.5$ and $q=0.1$. }
	\label{LT_SC}
	\end{center}
\end{figure}

\textbf{Low temperature with strong couplings ($T_{\rm{low}}, \beta_{\rm{small}} $):} Expanding the ADM mass about $\beta=0$, we obtain 
\begin{eqnarray}
  M|_{\rm{small}\, \beta}&=&\frac{{{\Sigma _{D - 2}}}}{{16\pi {{\left( {D - 1} \right)}^2}{r_+^3}}} \times\nonumber \\
 &&\Biggr\{- 8{\left( {D - 1} \right)^2}q{r_+^4}\beta  - \frac{2}{\sqrt \pi  } \left( {D - 1} \right)\,\Gamma \left( {\frac{1}{{2D - 4}}} \right)\Gamma \left( {\frac{{1 - D}}{{2D - 4}}} \right)q\beta {{\left( {\frac{q}{\beta }} \right)}^{\frac{1}{{D - 2}}}}r_+^3 \nonumber\\
&& + {r_+^D}\left( {\left( {D - 1} \right)\left( {2 + D\left( {D - 3} \right) + 16\pi P{r_+^2}} \right) + 8\left( {2D - 3} \right){\beta ^2}{r_+^2} - 8\left( {D - 1} \right){\beta ^2}{r_+^2}\ln \left( {\frac{{2{r_+^{D - 2}}\beta }}{q}} \right)} \right)\Biggr\}.
\end{eqnarray}
Replacing $  M|_{\beta=0}$ in the exact efficiency formula \ref{EEF}, the efficiency of the ideal cycle for a logarithmic $U(1)$ AdS black hole will be as

\begin{eqnarray}\label{eff0}
\eta|_{\rm{small}\, \beta}=\frac{G(r_1,r_2,P_1,P_3)}{H(r_1,r_2,P_1,P_3)},
\end{eqnarray}
  where
 \addtocounter{equation}{-1}
 \begin{subequations}
 \begin{align}
G({r_1},{r_2},{P_1},{P_3}) &= 16{\pi ^{3/2}}\left( {D - 1} \right)\left( {{P_1} - {P_3}} \right)(r_1 r_2)^2\left( { - {r_1^D}{r_2} + {r_1}{r_2^D}} \right),\notag\\
  H(r_1,r_2,P_1,P_3)&= \sqrt \pi  \Biggr\{ r_1^3r_2^D\left( {\left( {D - 1} \right)\left( {2 + D\left( {D - 3} \right) + 16\pi {P_1}r_2^2} \right) + 8\left( {2D - 3} \right){\beta ^2}r_2^2} \right)\notag\\
   &+ 8\left( {D - 1} \right){\beta ^2}{\left( {{r_1}{r_2}} \right)^2}\left( {r_1^D{r_2}\ln \left( {\frac{{2r_1^{D - 2}\beta }}{q}} \right) - {r_1}r_2^D\ln \left( {\frac{{2r_2^{D - 2}\beta }}{q}} \right)} \right)\notag\\
   &+ r_2^3\left( {8{{\left( {D - 1} \right)}^2}q\beta r_1^3\left( {{r_1} - {r_2}} \right) - r_1^D\left( {\left( {D - 1} \right)\left( {2 + D\left( {D - 3} \right) + 16\pi {P_1}r_1^2} \right) + 8\left( {2D - 3} \right){\beta ^2}r_1^2} \right)} \right)\Biggr\}. 
 \end{align}
\end{subequations}
Notice that the same is obtained for the expansion around $r \to 0$, meaning that we are approaching the point charge, no matter how much is the strength of the non-linearity coupling $\beta$.  $\eta$, $\eta/\eta_{C}$ and $\eta/\eta_{0}$ are displayed in the range of $10^{-6}<\beta<10^{-1}$ for scheme 1 in Fig. \ref{LT_SC}. $\eta$ and $\eta/\eta_{0}$ are increasing functions as they were in other regions whereas $\eta/\eta_{C}$ first increases unlike in Figs. \ref{scheme1_fig2}-\ref{LT_WC} and then decreases with respect to $\beta$. In the interval in which $\eta/\eta_{C}$ is increasing, $\eta$ and $\eta_{C}$ are both decreasing but the rate of decrease of $\eta_{C}$ is greater than $\eta$ and, as a result, $\eta/\eta_{C}$ increases.
\par

\begin{figure}
\begin{center}
	\epsfxsize=9 cm 
	\includegraphics[width=5.5 cm]{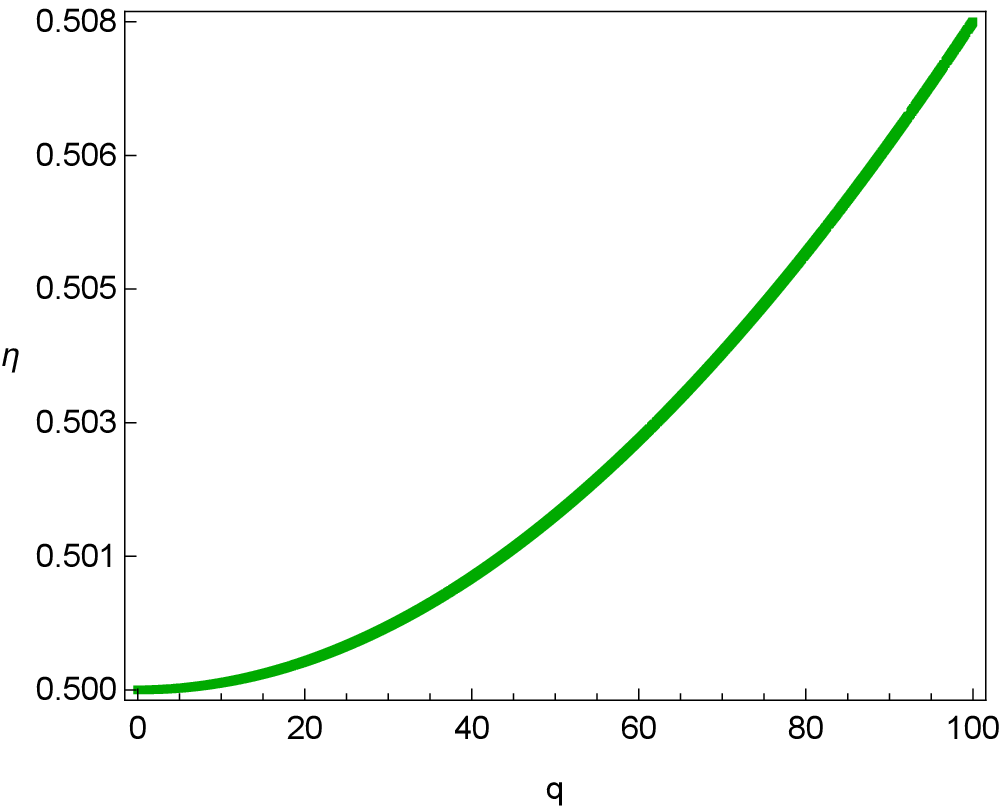}
	\hskip 0.5 cm
	\epsfxsize=9 cm 
	\includegraphics[width=5.5 cm]{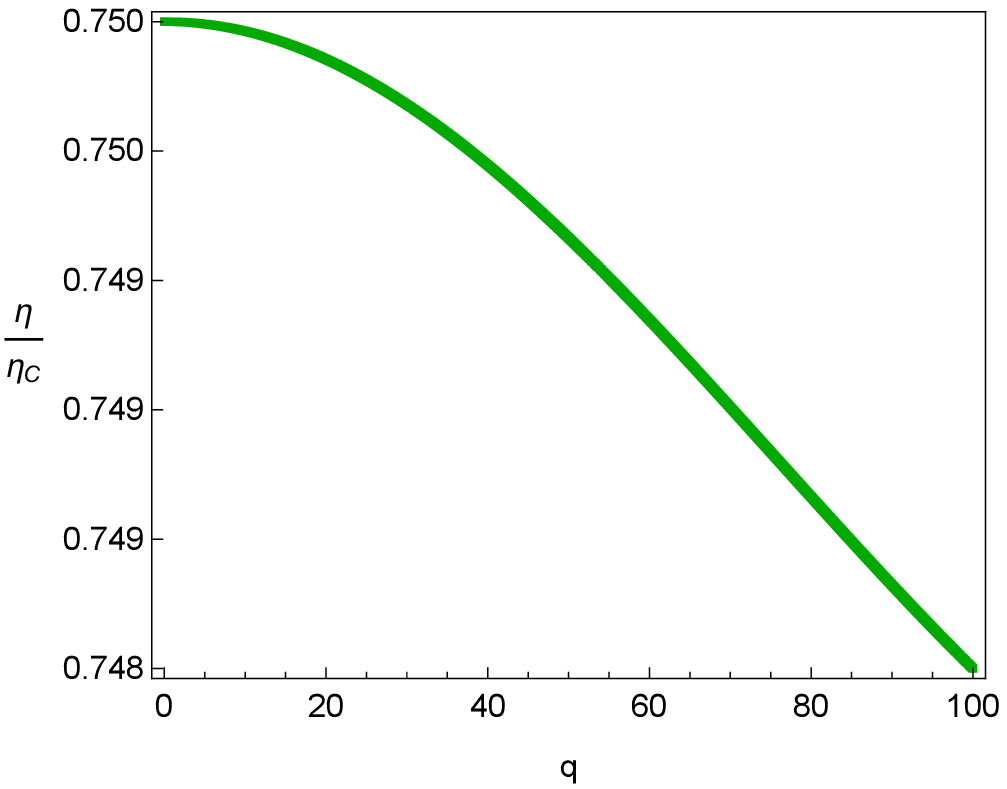}
	\hskip 0.5 cm
	\epsfxsize=9 cm 
    \includegraphics[width=5.5 cm]{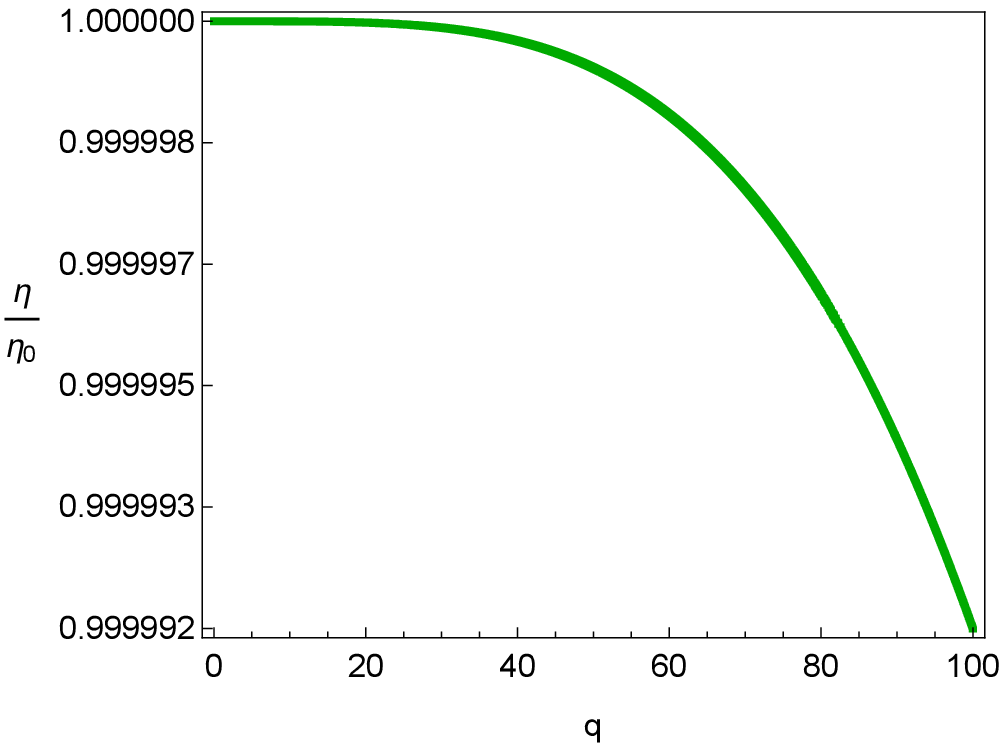}
	\caption{ $\eta$, $\eta/\eta_{C}$ and $\eta/\eta_{0}$ versus $q$ in scheme 1 for $D=4$, $T_1=20$, $T_2=30$, $P_1=1$, $P_4=0.5$ and $\beta=10$. }
	\label{scheme1_q}
	\end{center}
\end{figure}

\begin{figure}
\begin{center}
	\epsfxsize=9 cm 
	\includegraphics[width=5.5 cm]{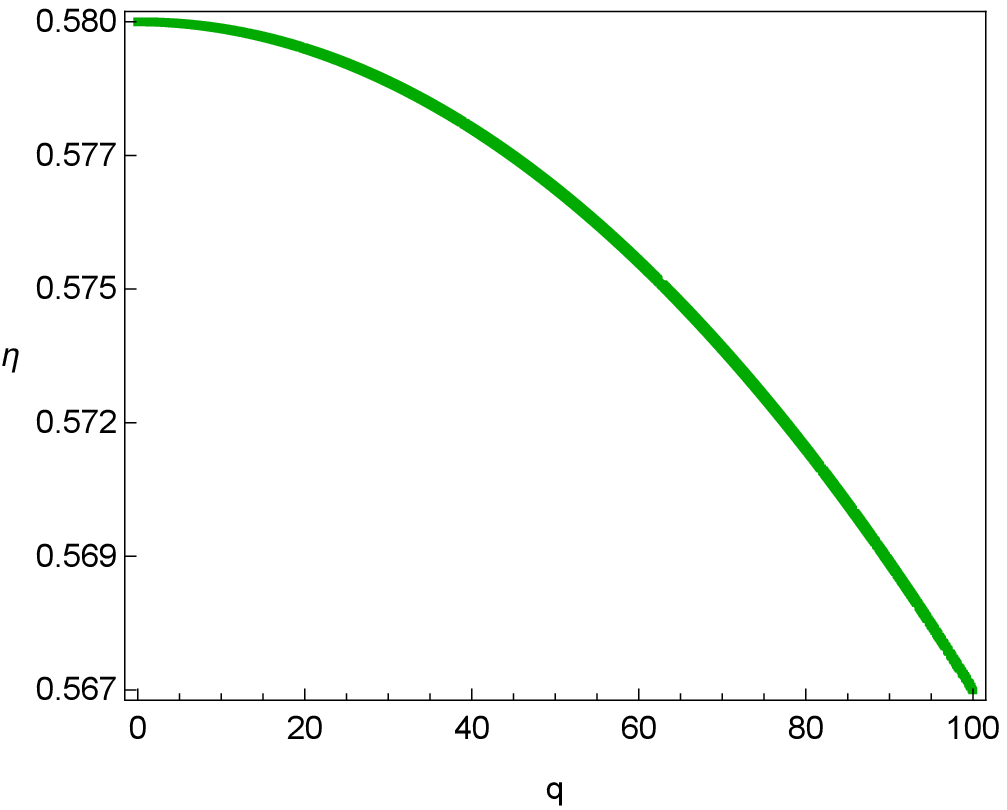}
	\hskip 0.5 cm
	\epsfxsize=9 cm 
	\includegraphics[width=5.5 cm]{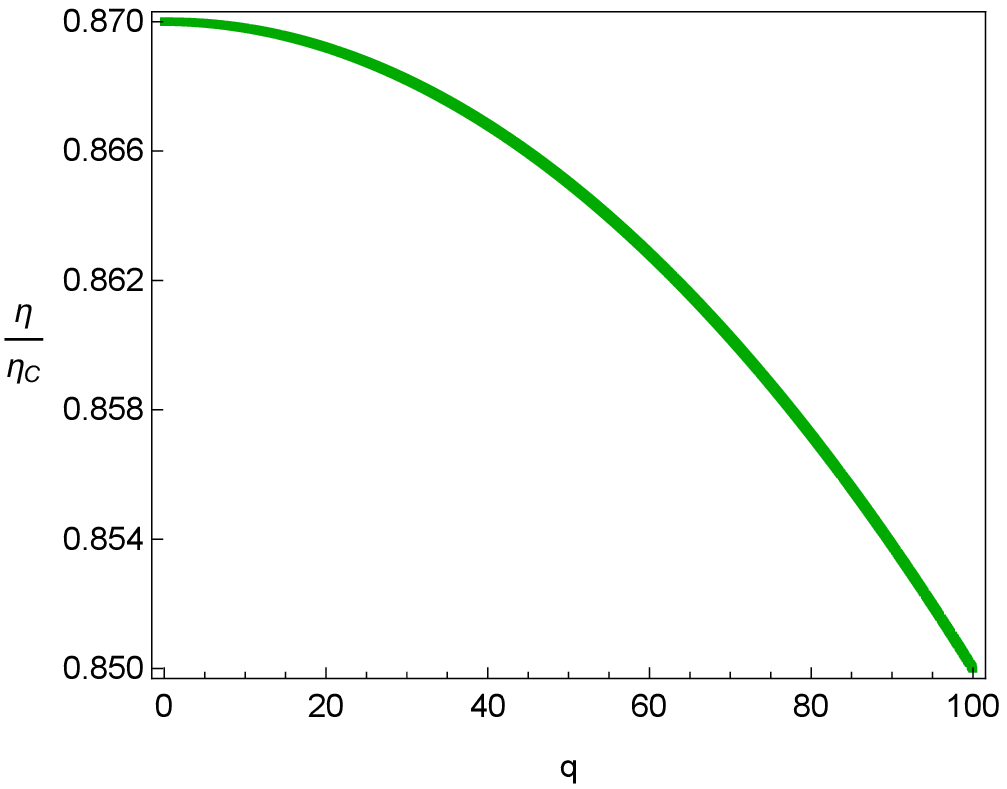}
	\hskip 0.5 cm
	\epsfxsize=9 cm 
    \includegraphics[width=5.5 cm]{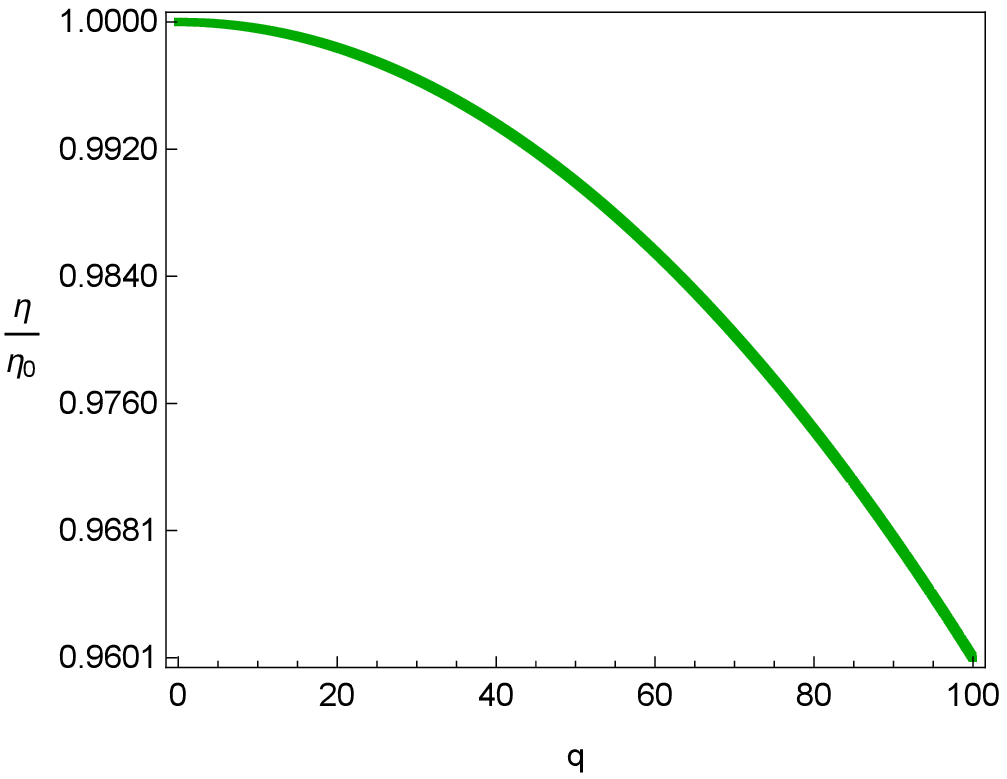}
	\caption{ $\eta$, $\eta/\eta_{C}$ and $\eta/\eta_{0}$ versus $q$ in scheme 2 for $D=4$, $V_2=10000$, $V_4=5000$, $T_2=30$, $T_4=10$ and $\beta=10$. }
	\label{scheme2_q}
	\end{center}
\end{figure}

From our physical intuition, we anticipate that in the high $U(1)$ charge limit, the metric function, mass, temperature and efficiency should behave in the same way as in the strong coupling limit. And in the low $U(1)$ charge limit, the same results are obtained as in the weak coupling limit. This can be easily  proved by expanding these quantities about $q=\infty$ (or equivalently $\beta=0$) and about $q=0$ (or equivalently $\beta=\infty$). These features can be seen in Figs. \ref{scheme1_q} and \ref{scheme2_q} where $\eta$, $\eta/\eta_{C}$ and $\eta/\eta_{0}$ are plotted versus $q$ in scheme 1 and scheme 2. As $q$ goes to zero, $\eta$ approaches $\eta_0$ in both schemes. 

\subsection{Efficiency for hyperbolic and planar black hole heat engines} \label{planar-hyperbolic engines}
\begin{figure}
\begin{center}
	\epsfxsize=9 cm 
	\includegraphics[width=5.5 cm]{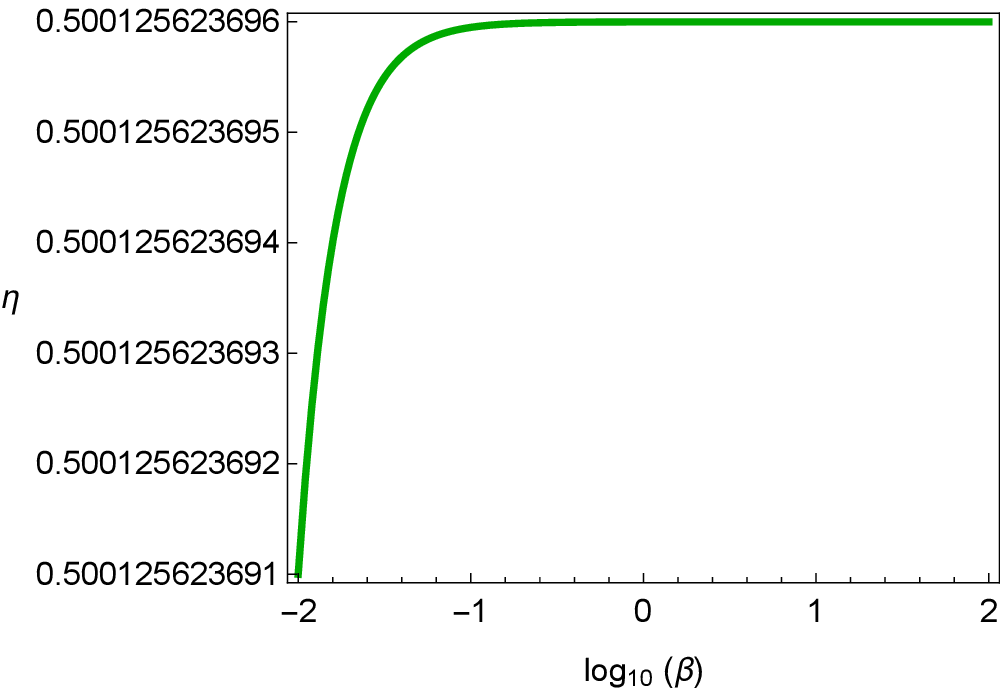}
	\hskip 0.5 cm
	\epsfxsize=9 cm 
	\includegraphics[width=5.5 cm]{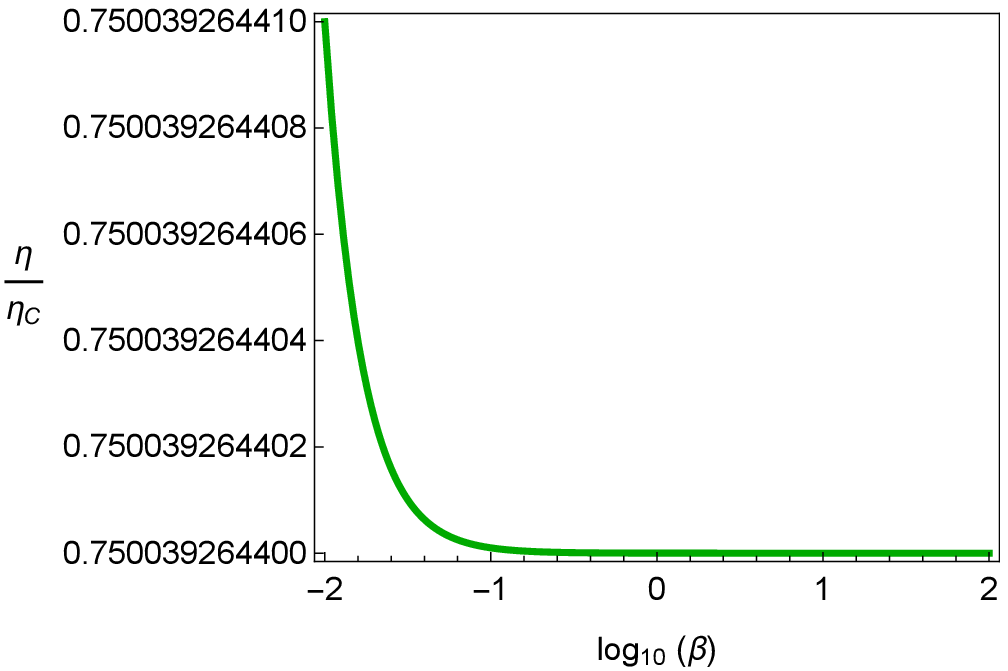}
	\hskip 0.5 cm
	\epsfxsize=9 cm 
    \includegraphics[width=5.5 cm]{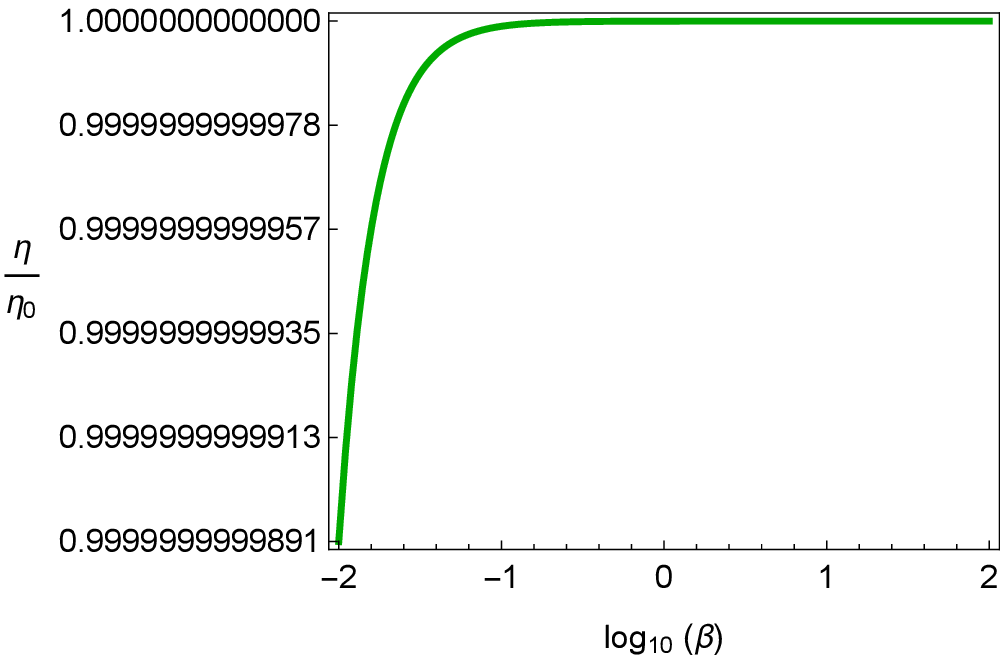}
	\caption{$\eta$, $\eta_{C}$ and $\eta/\eta_{0}$ versus log$_{10}(\beta)$ for hyperbolic black hole heat engines in scheme 1. Here we take$D=4$, $T_1=20$, $T_2=30$, $P_1=1$, $P_4=0.5$ and $q=0.1$ as in Fig. \ref{scheme1_fig2}. }
	\label{hyperbolic_scheme1}
	\end{center}
\end{figure}

\begin{figure}
\begin{center}
	\epsfxsize=9 cm 
	\includegraphics[width=5.5 cm]{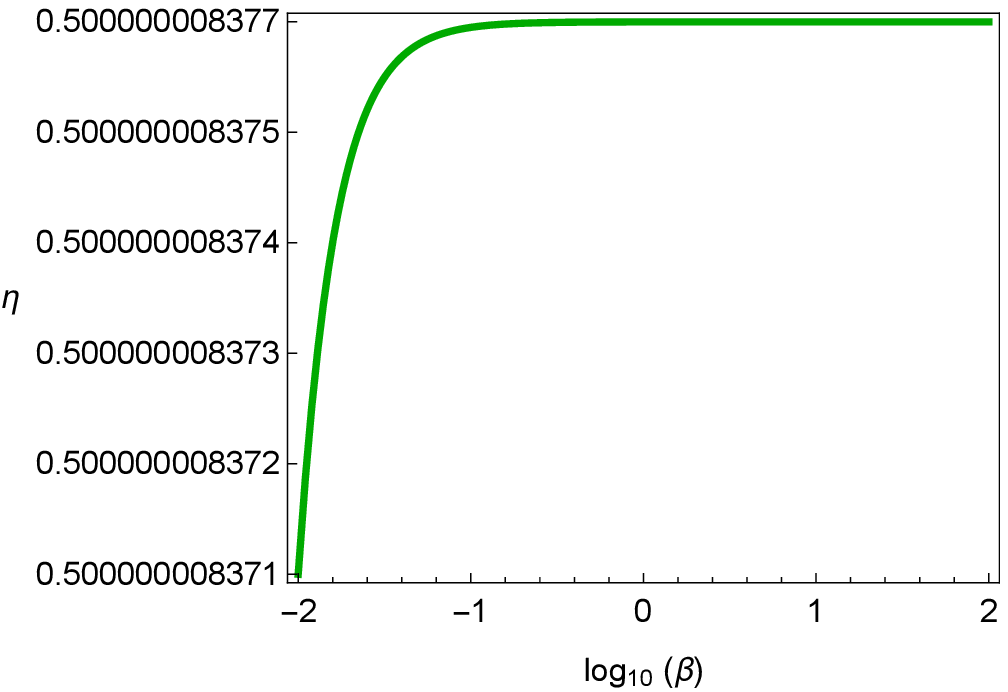}
	\hskip 0.5 cm
	\epsfxsize=9 cm 
	\includegraphics[width=5.5 cm]{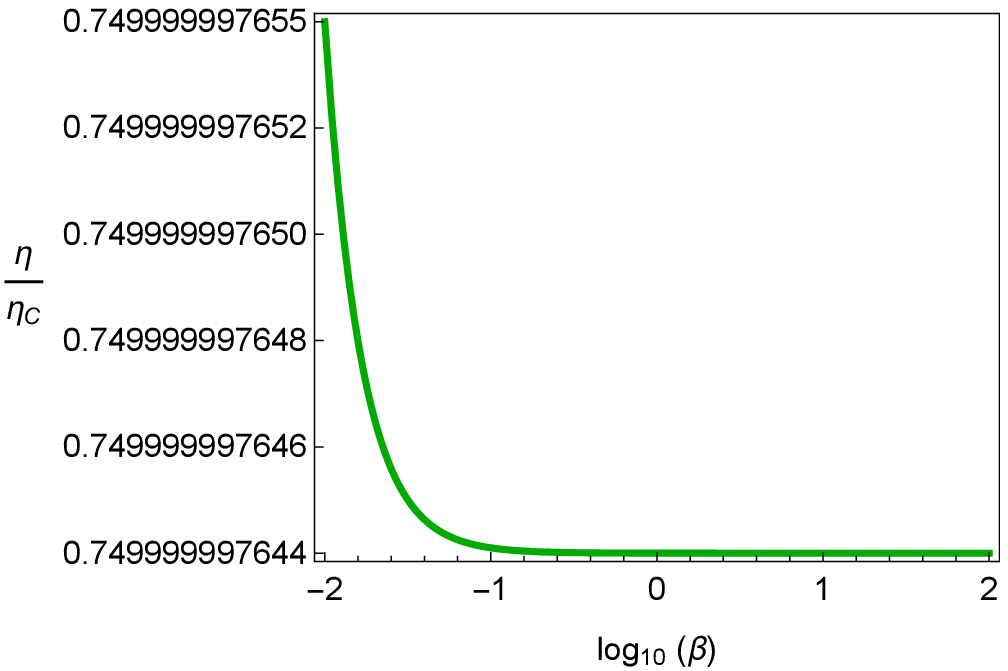}
	\hskip 0.5 cm
	\epsfxsize=9 cm 
    \includegraphics[width=5.5 cm]{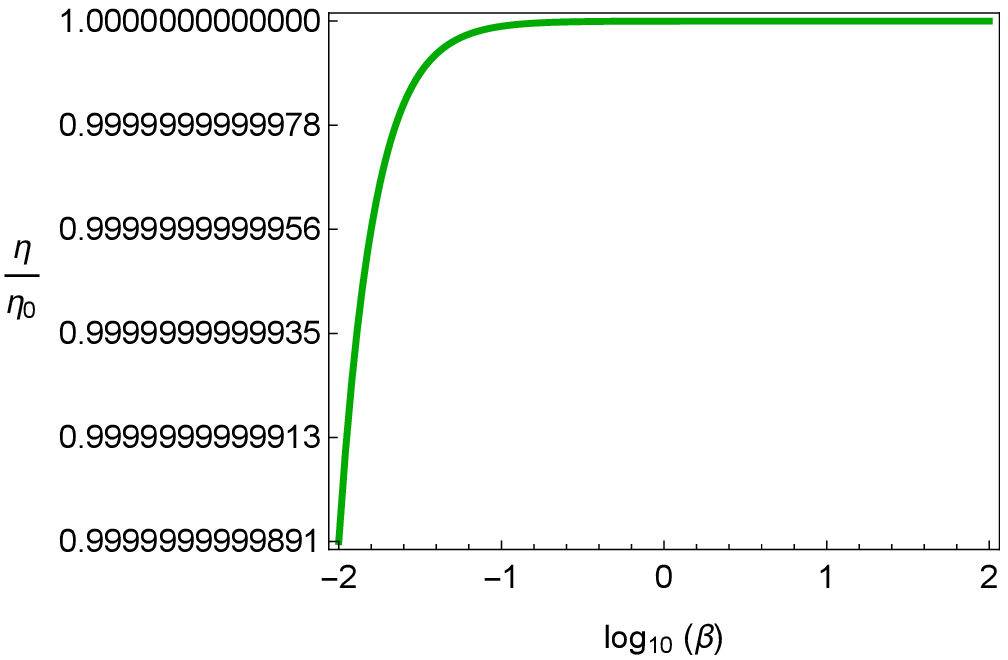}
	\caption{$\eta$, $\eta_{C}$ and $\eta/\eta_{0}$ versus log$_{10}(\beta)$ for planar black hole heat engines in scheme 1. Here we take $D=4$, $T_1=20$, $T_2=30$, $P_1=1$, $P_4=0.5$ and $q=0.1$ as in Fig. \ref{scheme1_fig2}.   }
	\label{planar_scheme1}
	\end{center}
\end{figure}

\begin{figure}
\begin{center}
	\epsfxsize=9 cm 
	\includegraphics[width=5.5 cm]{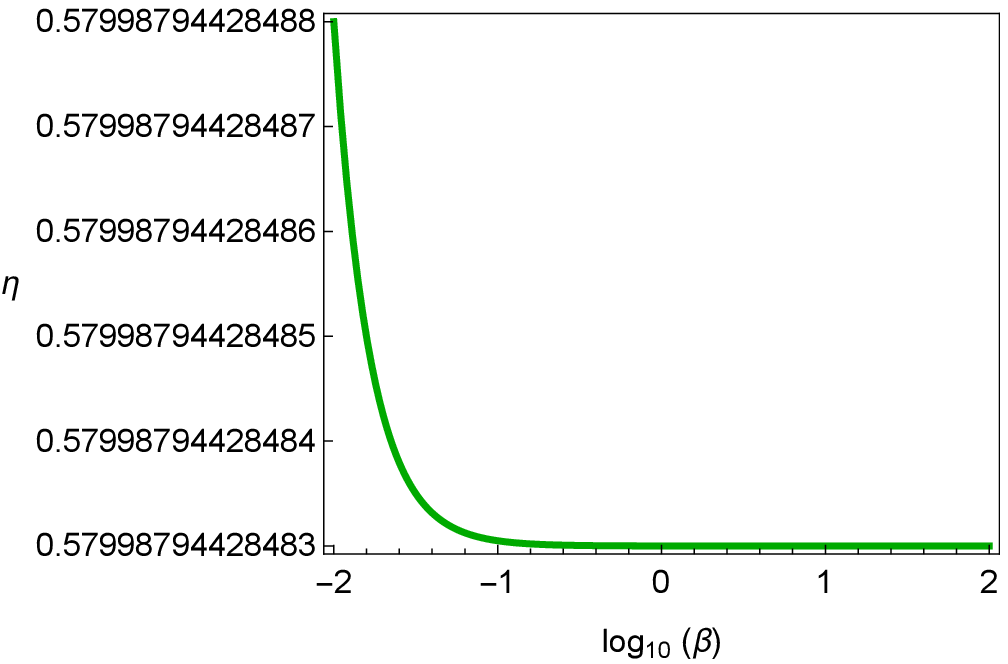}
	\hskip 0.5 cm
	\epsfxsize=9 cm 
	\includegraphics[width=5.5 cm]{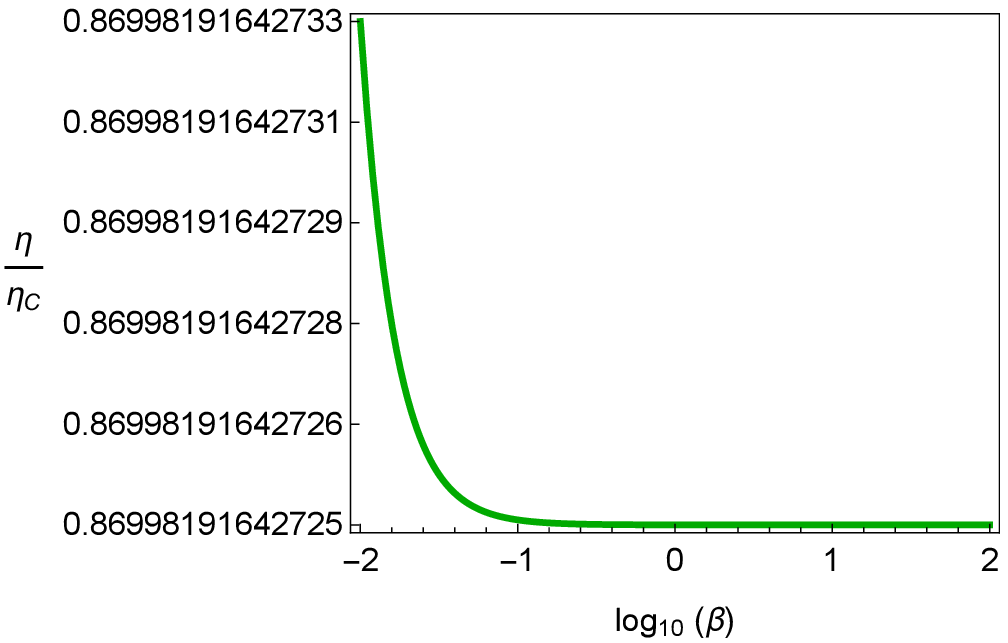}
	\hskip 0.5 cm
	\epsfxsize=9 cm 
    \includegraphics[width=5.5 cm]{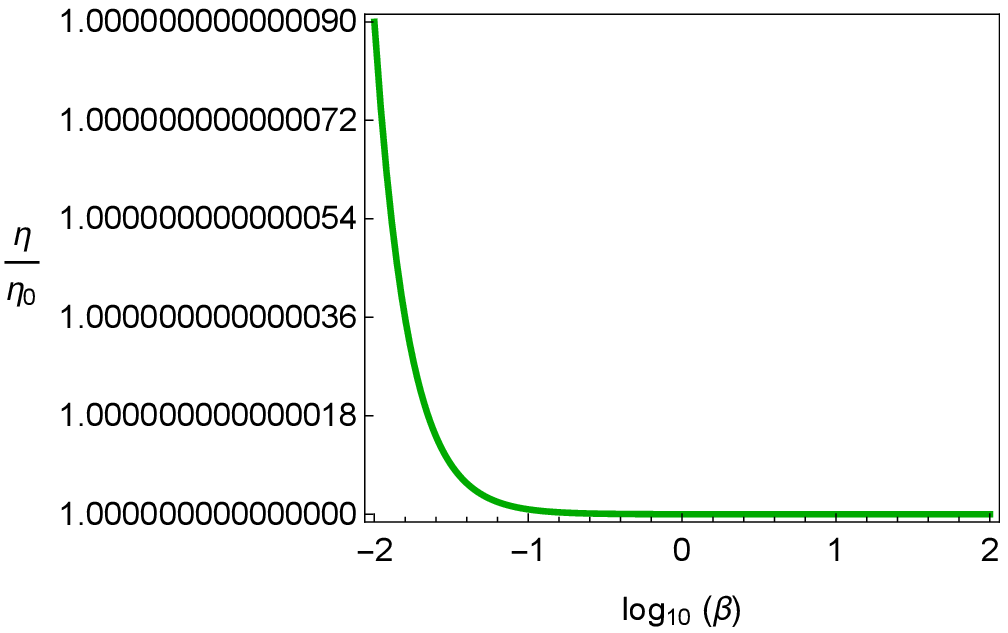}
	\caption{ $\eta$, $\eta/\eta_{C}$ and $\eta/\eta_{0}$ versus log$_{10}(\beta)$ for hyperbolic black hole heat engines in scheme 2. Here we take $D=4$, $T_2=30$, $T_4=10$, $V_2=10000$, $V_4=5000$ and $q=0.1$. }
	\label{hyperbolic_scheme2}
	\end{center}
\end{figure}

\begin{figure}
\begin{center}
	\epsfxsize=9 cm 
	\includegraphics[width=5.5 cm]{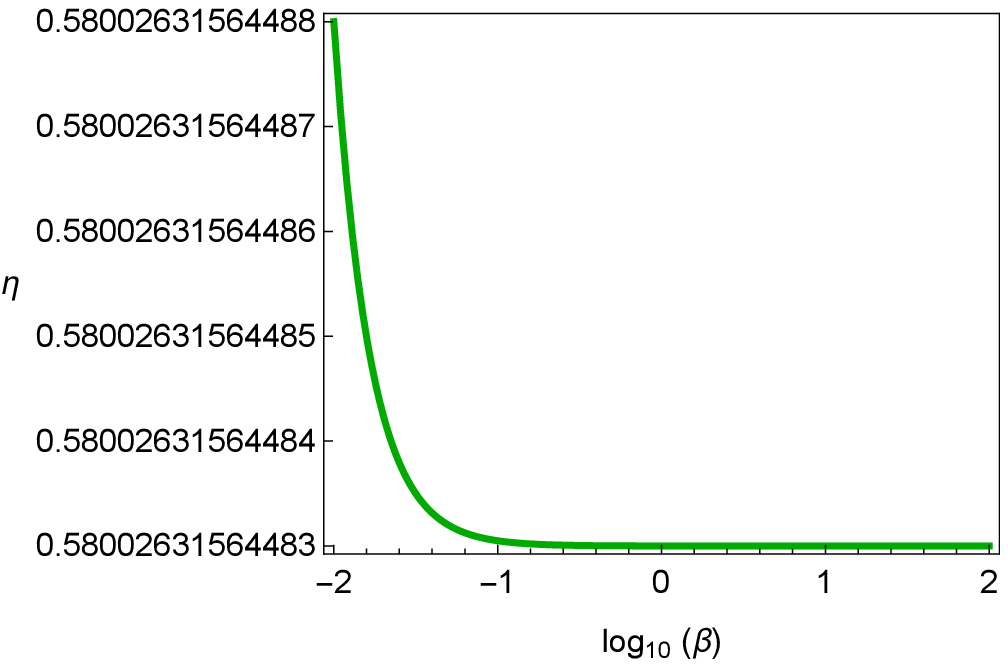}
	\hskip 0.5 cm
	\epsfxsize=9 cm 
	\includegraphics[width=5.5 cm]{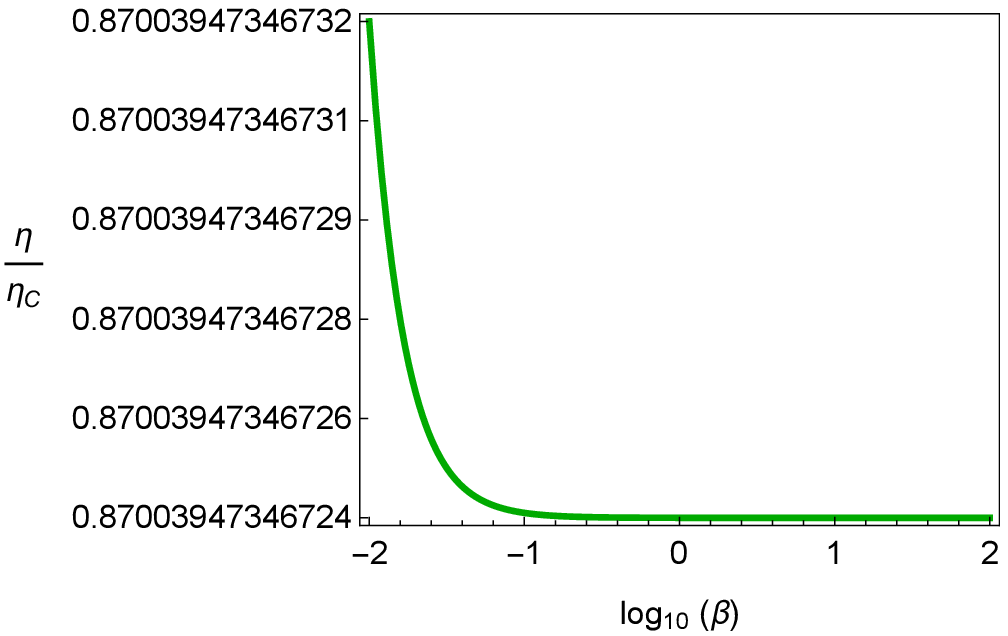}
	\hskip 0.5 cm
	\epsfxsize=9 cm 
    \includegraphics[width=5.5 cm]{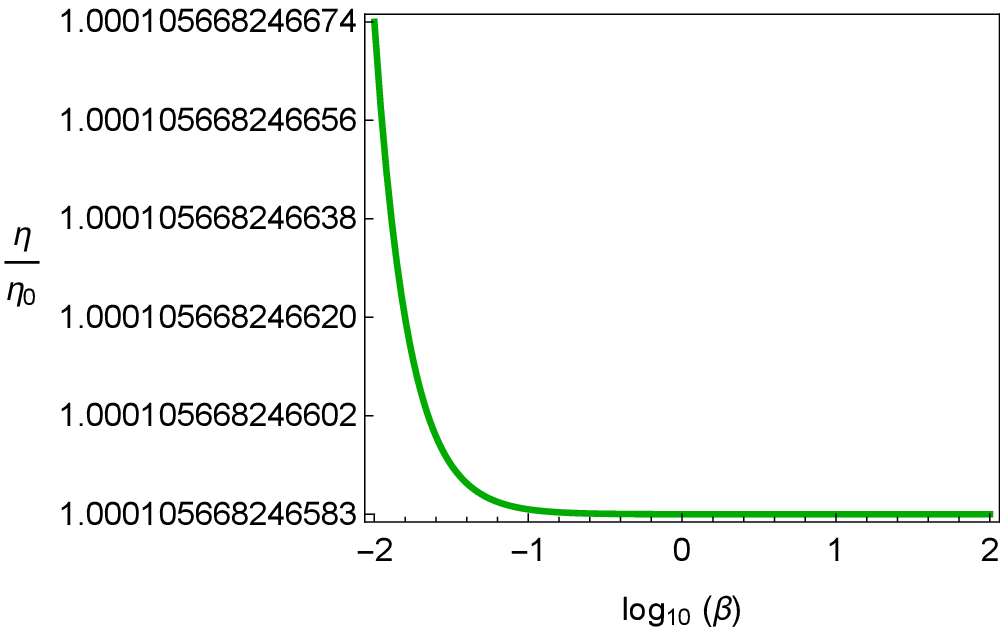}
	\caption{$\eta$, $\eta/\eta_{C}$ and $\eta/\eta_{0}$ versus log$_{10}(\beta)$ for planar black hole heat engines in scheme 2. Here we take $D=4$, $T_2=30$, $T_4=10$, $V_2=10000$, $V_4=5000$ and $q=0.1$.}
	\label{planar_scheme2}
	\end{center}
\end{figure}

Here, we concisely discuss the cases of planar and hyperbolic black holes as holographic heat engines. Apparently, the mass and therefore the efficiency of the black hole heat engine \ref{MADM} depend on the topology of the event horizon. Using the exact efficiency formula \ref{EEF}, we plot $\eta$, $\eta/\eta_{C}$ and $\eta/\eta_{0}$ versus log$_{10}(\beta)$ for black hole heat engines that are hyperbolic or planar in topology in Figs. \ref{hyperbolic_scheme1}-\ref{planar_scheme2} with the same parameters we have used in Fig. \ref{scheme1_fig2} for scheme 1 and \ref{scheme2_fig} for scheme 2.

 In scheme 1, similar to the spherical case, for both planar and hyperbolic black hole heat engines, $\eta$ and $\eta/\eta_{0}$ are increasing functions and $\eta/\eta_{C}$ is a decreasing function (see Figs \ref{hyperbolic_scheme1} and \ref{planar_scheme1}) and their behaviours are qualitatively the same as in Fig. \ref{scheme1_fig2}. Comparing Figs. \ref{scheme1_fig2}, \ref{hyperbolic_scheme1} and \ref{planar_scheme1} with each other, it is found that in this scheme the hyperbolic black hole heat engines are the most efficient and the spherical black hole heat engines work the least efficiently.
 
  Furthermore, in scheme 2, the same qualitative behaviour for $\eta$, $\eta/\eta_{C}$ and $\eta/\eta_{0}$  as in the spherical black hole heat engines is seen for planar and hyperbolic heat engines. By comparing Figs. \ref{scheme2_fig}, \ref{hyperbolic_scheme2} and \ref{planar_scheme2}, it is seen that in scheme 2,  the order is reversed. i.e., black hole heat engines with spherical symmetry have higher efficiencies than planar and hyperbolic.

\section{Closing remarks} \label{Final}
Logarithmic $U(1)$ gauge theory (proposed by Soleng \cite{NLBHSoleng1995}) has attracted lot of attention due to its relation to string effective actions in low energies. In addition, this nonlinear theory of Electrodynamics can represent the properties of both the Born-Infeld and the Euler-Heisenberg actions. We have briefly reviewed the geometric and thermodynamic properties of AdS black hole solutions coupled with logarithmic nonlinear electrodynamics. The non-linearity of $U(1)$ electromagnetic field, controlled by parameter $\beta$, affects the extended first law and the Smarr relation, which leads to new thermodynamic prospects. As highlighted in Sec. \ref{Thermodynamics}, the appearance of new pair ``${\cal B} \beta$'' in the Smarr relation is characteristic of BI-type nonlinear electrodynamics and is absent in linear Maxwell or power Maxwell invariant theories of Electrodynamics. We also found that the logarithmic vacuum polarization (${\cal B}$) at short and large distances from the charged source behaves as
\begin{equation}
{\cal B} \to \left\{ \begin{array}{l}
\mathop {\lim }\limits_{{r_ + } \to \infty } {\cal B} = 0\,\,\,\,\,\,\,\,\,\,{\rm{for \, any \, fixed }} \,\beta \\
\mathop {\lim }\limits_{{r_ + } \to 0} {\cal B} = {\rm{finite}}\,\,\,\,\,\,\,\,\,\,{\rm{for \, any \, fixed }}\, \beta 
\end{array} \right.
\end{equation}
while in strong ($\beta \to 0$) and weak ($\beta \to \infty$) coupling regimes, this quantity behaves as
\begin{equation}
{\cal B} \to \left\{ \begin{array}{l}
\mathop {\lim }\limits_{\beta  \to 0} {\cal B} = \infty \,\,\,\,\,\,\,\,\,\,{\rm{for \, any \, fixed }} \, {r_ + }\\
\mathop {\lim }\limits_{\beta  \to \infty } {\cal B} = {\rm{0}}\,\,\,\,\,\,\,\,\,\,{\rm{for \, any \, fixed }} \, {r_ + } \,.
\end{array} \right.
\end{equation}
In both the expansions around $\beta = \infty$ and $r_+ = \infty$, the vacuum polarization has similar form, indicating that the high temperature expansion ($T \propto r_+$ for large $r_+$) is equivalent to the weak coupling limit ($\beta \to \infty$), as analytically proved in Sect. \ref{Heat Engines}.

We considered the logarithmic $U(1)$ AdS black hole as a working substance of a holographic heat engine which undergoes a rectangular cyclic process consisting two isobars and two isochores/adiabats. Using the exact efficiency formula, we studied the efficiency of the spherical black hole heat engine as a function of $\beta$ in two schemes. These schemes differ by the choice of thermodynamic cycle parameters which are fixed under changing the non-linearity parameter $\beta$. We saw that efficiency changes by exceedingly small amounts with respect to $\beta$ for both strong and weak couplings, although the changes are more sensible in strong couplings. In general, the nonlinear corrections to Maxwell electrodynamics are so tiny, as expected. As an example, light-by-light scattering $(\gamma\gamma\rightarrow \gamma\gamma)$, which is a very rare phenomenon, can be classically described within the context of nonlinear electrodynamics theories \cite{HeisenbergEuler1936,JacksonElectrodynamics,SchwartzQFT} and quantum mechanically by radiative corrections in QED \cite{SchwartzQFT,Klauber}. After a long search, its evidence has been recently reported by the ATLAS Collaboration \cite{ATLAS}.
\par
 In order to compare our results with the Carnot efficiency, the maximum efficiency available by a heat engine working between two temperatures, we plotted $\eta/\eta_{C}$ for both schemes. It was found that $\eta/\eta_{C}$ decreases with increasing $\beta$, meaning that the holographic black hole heat engine works more efficiently with respect to the maximum possible efficiency in the strong coupling limit. In other words, the engine efficiency with respect to its possible maximum value increases when nonlinear effects of electrodynamics are taken into account, where, as a consistency check, it is in agreement with the results of \cite{Johnson2016a} and \cite{HeatEngines2017b}. Also, $\eta/\eta_{0}$ was displayed for different schemes and as expected it approaches to $1$ when $\beta$ goes to infinity. Exploring the low temperature domain, the same qualitative behaviour for $\eta/\eta_{C}$ and $\eta/\eta_{0}$  was seen. Furthermore, it was observed that in general efficiency varies too slowly in $\beta$ especially for $\beta>0.1$. 
\par 
Then we investigated the limiting behaviour of efficiency in the weak and strong coupling regimes and found the analytic relations for high/low temperature domains with weak/strong couplings. In the high temperature limit, we used the fact that $T \propto r_{+}$ and made series expansions of $Q_H$, $W$ and indeed $\eta$. The results show that for weak couplings ($\beta\rightarrow \infty$) in the high temperature domain, the change of efficiency with $\beta$ is imperceptible while it is sensible in the strong coupling regime ($\beta\rightarrow 0$). The similar qualitative behaviour of $\eta$ in weak and strong coupling regimes in the high temperature limit indicates that the efficiency is a smooth monotonic function of $\beta$ for $0<\beta<\infty$. In the low temperature limit with weak couplings, we made an expansion series around $\beta=\infty$ and ran it in the low working temperature, since expanding around $T=0$ is impossible. Again, $\eta$, $\eta/\eta_{C}$ and $\eta/\eta_{+}$ were explored in this domain and it was found that they behave the same way as in the high temperature domain. It was seen that for the strong couplings in the low temperature limit, high enough to be greater than critical temperature, $\eta/\eta_{C}$ versus log$_{10}(\beta)$ first increases and then decreases.
\par
Moreover, we studied the planar and hyperbolic black holes in the presence of logarithmic non-linear electrodynamics as holographic heat engines. Qualitatively, the same behaviour as in spherical holographic heat engines was seen for efficiency in both schemes. Quantitatively, hyperbolic black hole heat engines are the most efficient in scheme 1 but, in scheme 2, the spherical holographic heat engines works more efficiently. In summary, we observed
\begin{itemize}
	\item \textit{in scheme 1}:
\begin{equation}
{\eta _{{\rm{spherical}}}} < {\eta _{{\rm{planar}}}} < {\eta _{{\rm{hyperbolic}}}}
\end{equation}
\end{itemize}
\begin{itemize}
	\item \textit{in scheme 2}:
\begin{equation}
{\eta _{{\rm{hyperbolic}}}} < {\eta _{{\rm{planar}}}} < {\eta _{{\rm{spherical}}}}
\end{equation}
\end{itemize}
These relations are also hold for the ratio of efficiency with Carnot efficiency ($\eta/\eta_C$). As it is clear, the results are scheme/parameter dependent. The reason is that the equation of state depends on $\beta$ and as a result the parameters of the cycle, which are not hold fixed, depend on  $\beta$. This leads to the opposite behaviour for the efficiency of black holes with different topologies in the mentioned schemes.
\par 
Finally, it should be noted that Logarithmic $U(1)$ charged AdS black holes can be understood more naturally in the context of the extended phase space thermodynamics. However, it is not clear that what kinds of thermodynamic phase transitions are allowed in these black holes and it would be interesting to investigate about their critical phenomena. In addition, the Joule-Thomson expansion for these black hole solutions can be studied by isenthalpic and inversion curves in $T-P$ plane in order to explore about the effect of nonlinear electromagnetic corrections to this process.

\begin{acknowledgements}
We gratefully thank the anonymous referee for enlightening comments and suggestions which substantially helped in improving the quality of the paper. S.Z. would like to express her sincere gratitude to A. Dehghani for useful discussions. S.Z. also appreciates the support of University of Sistan and Baluchestan research council.
\end{acknowledgements}


\end{document}